%% file: dust6.tex
\documentclass[useAMS,usenatbib]{my_mn2e}
\usepackage{color,graphicx}
\usepackage{moreverb}
\usepackage{times}

\usepackage{float,cancel,rotating}

\title[All-sky Observational Evidence for A Temperature Dependent Emissivity Spectral Index]{All-sky Observational Evidence for An Inverse Correlation between Dust Temperature and Emissivity Spectral Index}
\author[Z.~Liang, D. J. Fixsen and B. Gold]{Z.~Liang,$^1$\thanks{E-mail: zliang1@jhu.edu} D.~J.~Fixsen$^{2}$ and B.~Gold$^{1}$ \\
$^1$Department of Physics \& Astronomy, The Johns Hopkins University, 3400 N. Charles St., Baltimore, MD, 21218, USA\\
$^2$University of Maryland, Goddard Space Flight Center, MD, 20771, USA}

\begin{document}

\pagerange{\pageref{firstpage}--\pageref{lastpage}} \pubyear{2011}

\maketitle

\label{firstpage}

\begin{abstract}

We show that a one-component variable-emissivity-spectral-index model (the free-$\beta$ model) provides more physically motivated estimates of dust temperature at the Galactic polar caps than one- or two-component fixed-emissivity-spectral-index models (fixed-$\beta$ models) for interstellar dust thermal emission at far-infrared and millimeter wavelengths. For the comparison we have fit all-sky one-component dust models with fixed or variable emissivity spectral index to a new and improved version of the 210-channel dust spectra from the {\sl COBE}-FIRAS, the $100-240$~$\mu$m maps from the {\sl COBE}-DIRBE, and the 94~GHz dust map from the {\sl WMAP}. The best model, the free-$\beta$ model, is well constrained by data at $60-3000$~GHz over 86~per~cent of the total sky area. It predicts dust temperature ($T_\mathrm{dust}$) to be $13.7-22.7$~($\pm 1.3$)~K, the emissivity spectral index ($\beta$) to be $1.2-3.1$~($\pm 0.3$), and the optical depth ($\tau$) to range $0.6-46 \times 10^{-5}$ at $\nu_0 = 900$~GHz ($\lambda_0 = 333 \mu$m) with a 23~per~cent uncertainty. Using these estimates, we present all-sky evidence for an inverse correlation between the emissivity spectral index and dust temperature, which fits the relation $\beta = 1/(\delta+\omega \cdot T_\mathrm{dust})$ with $\delta = -0.510 \pm 0.011$ and $\omega = 0.059 \pm 0.001$. This best model will be useful to cosmic microwave background experiments for removing foreground dust contamination and it can serve as an all-sky extended-frequency reference for future higher resolution dust models.

\end{abstract}


\section{Introduction}
\input{intro_dust6}

\section[]{Observations}\label{sec:instrument_data}
\input{instruments_data_dust6}

\section[]{Data Preparation}\label{sec:data_preparation}
\input{data_preparation_dust6}

\section{Results and Analysis}\label{sec:results}
\input{analysis4ps_dust6}
\input{one_comp_var_res_dust6}

\section{Discussion}\label{sec:comparison}
\input{fixed_vs_free_alpha_dust6}

\input{discussion_dust6}

\section{Conclusions}\label{sec:conclusion}
\input{conclusion_dust6}

\section*{Acknowledgments}

Some of the results in this paper have been derived using the HEALPix (K.M. G\'{o}rski et al., 2005, ApJ, 622, p759) package.

We thank the anonymous reviewer for many invaluable suggestions which help improve the quality of this manuscript. 
ZL thanks Professor Charles Bennett for enlightening discussions, intuitive guidance and generous support, all of which made this work possible. ZL also thanks Drs. Janet Weiland, David Larson and Domenico Tocchini-Valentini for very helpful discussions.  

\input{bibliography_dust6}

\bsp

\label{lastpage}

\end{document}

%% file: intro_dust6.tex
An accurate model of thermal dust emission at the far-infrared and millimetre wavelengths is important for cosmic microwave background (CMB) anisotropy studies because it helps to remove one of the three major diffuse foreground contaminants. In the last decade, experiments such as the {\sl Wilkinson Microwave Anisotropy Probe} ({\sl WMAP}, \citealt{Bennett03a}) have precisely measured the angular variations in CMB signal in order to understand the global geometry and expansion of the universe. However, studying variations that are {\bf $10^{-5}$} the strength of the principal signal is difficult, and the removal of contaminating signals in the data needs to be done accurately. For these experiments, a dust template, such as one extrapolated from the \citet*{Finkbeiner99} (FDS) study of interstellar dust in the far-infrared has been used to remove thermal dust contribution from sky measurements \citep{Bennett03b, Hinshaw07, Gold09}.  

Among major efforts to derive an all-sky dust model from observational data, \citet{Reach95} use dust spectra derived from measurements of the Far Infrared Absolute Spectrophotometer (FIRAS, \citet{FIRAS97}) instrument on board the {\sl Cosmic Background Explorer} ({\sl COBE}, \citealt{Mather82}) satellite to constrain dust models with emissivity proportional to $\nu^2$. They found that dust emission is best described by a three-component dust model: a warm ($16-21$~K) and a cold ($4-7$~K) component that are present everywhere in the sky, and an intermediate temperature ($10-14$~K) component that exists only at the Inner Galaxy. 
In 1996, Boulanger et al. independently derived another set of dust spectra using the FIRAS measurements and fit to it a one-component $\nu^2$ emissivity dust model. They found that the average spectrum of dust associated with HI gas had an averaged temperature of $17.5 \pm 0.2$~K. In 1998, Lagache et al. used DIRBE bands at 100, 140 and 240~$\mu$m to decompose FIRAS spectra at $|b|>10^\circ$ into a cirrus and a cold component. For 61~per~cent of the sky where the cold emission was negligible, they found that the cirrus had a mean temperature of 17.5~K with a dispersion of 2.5~K. For the 3.4~per~cent sky where both cirrus and cold components were present, the two components were both assumed to follow a $\nu^2$ emissivity law, with the cirrus component found to have a temperature of $17.8 \pm 1.2$~K, and the cold component with a temperature of $15 \pm 0.8$~K.
A widely used dust model in CMB studies was obtained by \citet{Finkbeiner99}. Their best model (Model \#8) to the FIRAS dust spectra consisted of two components: a cold component following a $\nu^{1.67}$ emissivity law with temperature at $7.7-13.1$~K, and a warm component following a $\nu^{2.70}$ emissivity law with temperature at $13.6-21.2$~K. A decade later, the {\sl Planck} Collaboration \citep{Planck11a} used the {\sl Planck}-HFI (350~$\mu$m--2mm) and IRAS 100~$\mu$m data to derive all-sky dust temperature and optical depth maps using a one-component model with emissivity proportional to $\nu^{1.8}$. They found that the median temperature of the sky at 10$^\circ$ above and below the Galactic plane was 17.7~K, see also \citet{Liang11}. 

The above results highlight the diverse findings in the study of thermal dust emission at far-infrared and millimetre wavelengths. They show that the derived dust properties depend as much on the fitting method and the functional form of the model as on the data. With added new and more sensitive data from the {\sl WMAP}, we now can constrain model parameters with much greater accuracy.

A second reason for our work is to present fit results from applying variable spatial averaging to increase signal-to-noise of the spectra. Since low intensity measurements often come with large uncertainties, when such data are used directly to constraint a model, results are highly uncertain parameters. At times this problem is treated with averaging data within a predefined sky region. This approach has the disadvantage of using a presupposed dust distribution in the derivation of a solution while figuring out the distribution is part of the research question. Here we make no assumption of the dust distribution but instead use the signal-to-noise of the data to determine the amount of spatial averaging needed for the data. The results are higher spatial resolution for regions with good signal-to-noise and less averaging for the original data set. %This technique could be applied to similar problems in other areas of research. 

A third motivation for our work is to understand whether dust optical properties \citep{Draine84} are the same at far-infrared and millimetre wavelengths from the perspective of empirical model fitting. That far-infrared dust emissivity follows a $\nu^2$ power law has been widely accepted (see list above), yet the validity of such an extrapolation has not been proven by theory, laboratory experiment, or empirical model fitting. In fact, reports of laboratory measurements by \citet{Agladze96, Mennella98, Boudet05,Coupeaud11} have shown that emissivity of amorphous silicate and carbon grains differed from a $\nu^2$ power law and had a significant temperature dependence. This inconsistency between laboratory measurements and modeling of astronomical observations means that our understanding of dust emission in the far-infrared and millimetre is incomplete. In this work, we attempt to fill this gap by first deriving best-fitting dust models with emissivity spectral index fixed at different values and as a variable, and then comparing the quality-of-fit of these models. Based on our findings, we argue that dust emissivity differs in the far-infrared and millimetre from the optical.

Finally, regarding the many empirical models we now know, e.g. the ones listed above, one cannot help but ask: How do we test the validity of these models? Beside having good constraints on model parameters, are there physically motivated tests we can use to verify predictions of these models? Here, we propose one: to compare dust temperature distribution with the distribution of dust heating source at the Galactic poles. We conduct an independent and comprehensive test on all-sky one-component models, and show that all but the one-component free-$\beta$ model fail this test. 

The structure of our manuscript is as follows. In Section \ref{sec:instrument_data}, we review observations by the {\sl COBE} satellite's DIRBE and FIRAS experiments and the {\sl WMAP} satellite that are used in our model construction. In Section \ref{sec:data_preparation} we detail procedures taken to deduce a new set of FIRAS dust spectra and to unify calibrations of the data sets. In Section \ref{sec:results}, we present results and analysis from fitting one-component dust models with fixed and variable emissivity spectral index to the data. In Section \ref{sec:comparison}, we compare temperature predictions of the free-$\beta$ model with those of the fixed-$\beta$ models and show that only the free-$\beta$ model gives physically motivated predictions of dust temperature at the Galactic polar caps. We also discuss the implications of the free-$\beta$ model on the inverse correlation between emissivity spectral index and dust temperature. Conclusions along with suggestions for how to use our results are presented in Section \ref{sec:conclusion}.

%% file: instruments_data_dust6.tex
%The following sections review the instruments and the data sets we used in the following analysis.

\subsection{\textbf{\emph{COBE}} DIRBE}

The DIRBE instrument was a cryogenically cooled 10-band absolute photometer designed to measure the spectral and angular distribution of the diffuse infrared background. It had a 0$\fdg$7 beam and covered the wavelength range from 1.25 to 240~$\mu$m. During its lifetime, the DIRBE achieved a sensitivity of 10$^{-9}$~W~m$^{-2}$~sr$^{-1}$ at most wavelengths \citep{Boggess92,Silverberg93,DIRBE98}. 

We use the 1997 ``Pass 3b'' Zodi-Subtracted Mission Average (ZSMA) Maps at bands 100, 140, and 240~$\mu$m. These maps measure the Galactic and extragalactic diffuse infrared emission and have been calibrated to remove zodiacal light (zodi). They are available at the Legacy Archive for Microwave Background Data Analysis (LAMBDA)\footnote{The LAMBDA Web site is http://www.lambda.gsfc.nasa.gov/}.

\subsection{\textbf{\emph{COBE}} FIRAS}

The FIRAS instrument was a polarizing Michelson interferometer designed to precisely measure the difference between the CMB and a blackbody spectrum. The FIRAS had a 7$^\circ$ beam and covered the frequency range from $1-97$~cm$^{-1}$ at 0.45~cm$^{-1}$ resolution
\citep{Boggess92,Fixsen94a,FIRAS97}. 

We derive a new set of dust spectral maps from the Destriped Sky Spectra of the Pass 4 final data release using procedures described in Section \ref{sec:data_preparation}. The 210 6063-pixel maps comprise the main body of spectral information in our model fitting. 

Six types of uncertainties have been characterized by the FIRAS Team. The instructions on how to treat each of them are documented in \citet{Fixsen94,FIRAS97,Mather99}. Since we want to build models that respond to both spectral and spatial variations of dust emission, our analysis includes all six FIRAS uncertainties: 
detector noise (D), emissivity gain uncertainties (PEP), bolometer parameter gain uncertainties (JCJ), internal calibrator temperature errors (PUP), absolute temperature errors (PTP), and destriper errors ($\beta$). Section 7.10 of \citet{FIRAS97} provides very helpful instructions on how to assemble the covariance matrix. For example, the D and $\beta$ matrices vary only among pixels, while the PEP, JCJ, PUP and PTP matrices differ for different frequencies. Interested readers are referred to the \citet{FIRAS97} for details.

\subsection{\textbf{\emph{WMAP}}}

The Wilkinson Microwave Anisotropy Probe was designed to determine the geometry, content and evolution of the universe by measuring temperature anisotropy of the CMB radiation. It consisted of two back-to-back offset Gregorian telescopes and used 20 high electron mobility transistor (HEMT) based differential radiometers to measure the brightness difference between two lines of sight that were 141$^\circ$ apart. At five frequency bands: 23, 33, 41, 61, and 94~GHz, the {\sl WMAP} made full sky measurements, which were analyzed by the data processing pipeline and formed 13$\arcmin$ FWHM HEALPix\footnote{For definition and applications of the HEALPix projection, refer to \citet*{Gorski99}, \citet*{Gorski05}, \citet*{Calabretta07} and http://healpix.jpl.nasa.gov.} pixelization maps. The spin motion of the observatory and its scanning strategy symmetrized the {\sl WMAP} beams. Beam sizes were estimated using square-root of the beam solid angle. In the order of increasing frequencies, they were: 0.88, 0.66, 0.51, 0.35, and 0.22$^\circ$ \citep{Jarosik03, Page03, Hinshaw03,Jarosik07,WMAP08,Hinshaw09,Hill09}.  

We use the dust temperature map (at 94 GHz) derived from the ``base model'' in {\sl WMAP}'s Five-Year foreground modeling analysis by \citet{Gold09}. In the same study, Gold et al. used different models to account for the diffuse foreground emission at different {\sl WMAP} bands, with nonthermal synchrotron, thermal bremsstrahlung, and thermal dust as the standard components and tested the possible existence of steepening synchrotron and/or spinning dust. Their likelihood analysis has shown that the basic model with just three main foreground components would be sufficient to remove foregrounds from sky maps at high Galactic latitudes. 

\begin{table}
\centering
\caption{Spectral Coverage of DIRBE, FIRAS \& {\sl WMAP}}
\begin{tabular}{|l||c|c|c|} 
\hline
 & $\lambda$ & 1/$\lambda$ & $\nu$ \\ 
 & ($\mu$m) & (cm$^{-1}$) & (GHz) \\
\hline
DIRBE & ~100 & ~100 & 2998 \\
 & ~140 & ~~~71 & ~~42 \\
 & ~240 & ~~~42 & 1249 \\
FIRAS & $103-4407$ & $2-97$ & $68-2911$ \\
{\sl WMAP} & 3189 & ~~~~3 & ~~94 \\
\hline
\end{tabular}
\label{tbl:compare_bands}
\end{table}

%% file: data_preparation_dust6.tex
Since our objective has been to model both spectral and spatial variations of dust emission, we unified different hardware constraints and calibration standards to ensure that different data sets are compared on an equal footing. In Section \ref{sec:makefirdust} we explain the deduction of FIRAS dust spectra. In Sections \ref{sec:beam_difference} -- \ref{sec:temp_intensity} we explain treatments on the DIRBE and WMAP data to make sure that they have the same physical attributes as the FIRAS dust maps. Among procedures presented in this section, those pertaining to beam differences, map projections, spatial resolutions, DIRBE-FIRAS absolute calibrations, and temperature-flux conversion are applied to both signal and noise maps. Procedures on zodi zero-point corrections and FIRAS systematic errors are applied to noise maps only. The uncertainties used in our analysis are the quadratic sums of the uncertainties of data sets and models described in this section. More specifically, when we are given uncertainties of the parameters, we use 

\begin{displaymath}
V(I_i) = \sum_j \left( \frac{\partial I_i}{\partial x_j} \right)^2V(x_j)+ \sum_j \sum_{k\not= j}\left(\frac{\partial I_i}{\partial x_j}\right)\left(\frac{\partial I_i}{\partial x_k}\right) cov(x_j,x_k) \,
\end{displaymath}
to generate uncertainties of intensity predictions.

\subsection{Deducing FIRAS dust spectral maps} \label{sec:makefirdust}

\begin{figure}
	\centering
  \includegraphics[scale=0.33]{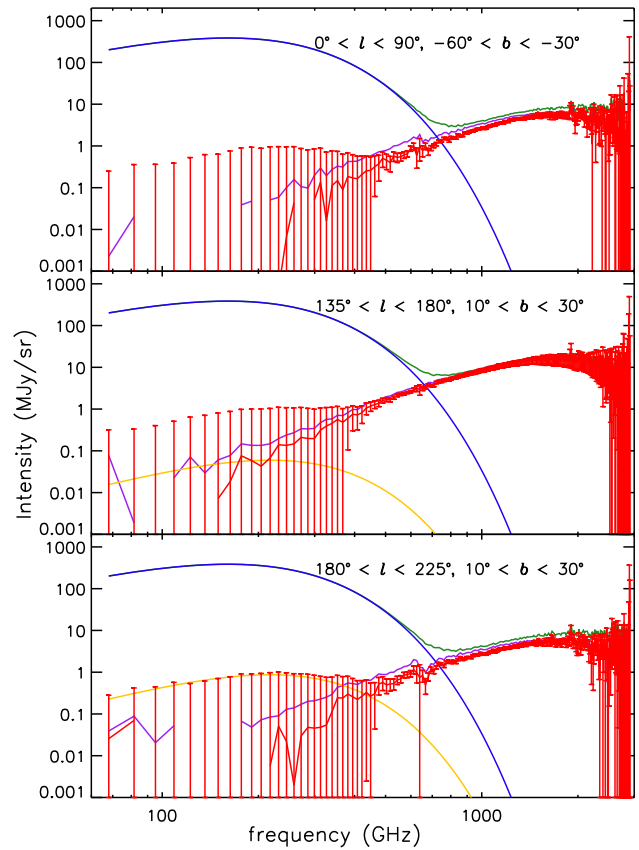}
	\caption{Examples of the new FIRAS dust spectra. Plotted in red are the new dust spectra and their error estimates; in purple are the dust spectra derived by the FIRAS team; in dark green is the total sky intensity measured by FIRAS; in blue is the CMB monopole; and in orange is the CMB dipole.} 
	\label{fig:new_FIRAS_dust}

\end{figure}

The FIRAS dust spectra on LAMBDA exhibit a ``jump'' between the low- and high-band data due to an inconsistent CMB monopole temperature subtraction. In addition, the Pass 4 data were released with an early calibration, making it necessary that we derive a new set of dust spectra. 
In the following we describe procedures to subtract from the Destriped Sky Spectra a blackbody spectrum for the CMB, a dipole of the Earth's motion with respect to the CMB, a zodi model, and contribution from the cosmic infrared background (CIB). Examples of the new dust spectra are plotted in red in Fig. \ref{fig:new_FIRAS_dust}. For comparison, corresponding dust spectra provided by the FIRAS Team are plotted in purple.

\subsubsection{CMB monopole and dipole}

The CMB temperature has been extensively treated in \citet{Mather90,Fixsen94,Mather94,Fixsen96,Mather99, Fixsen02,Fixsen09}. The appropriate correction for the Pass 4 data set is a 2.7278 black body spectrum.

A {\sl WMAP}-determined dipole \citep{Hinshaw09} has been removed from the Destriped Spectra. Specifically, $T_\mathrm{dipole} = 3.355$~mK and $(l,b)=(263\fdg 99, 48\fdg 26)$. Higher order variations in the CMB temperature were ignored because they were insignificant for this study.

\subsubsection{Zodi}

Zodi is the thermal emission and scattered light from interplanetary dust in our solar system. \citet{Kelsall98} derived a time-dependent parametric model for its emission, and the FIRAS Team extended those results to the entire FIRAS frequency coverage \citep{FIRAS97,Fixsen02}. Derivation of the zodi model for FIRAS hinged on the fact that FIRAS measurements overlap with DIRBE bands at 140 and 240~$\mu$m. Therefore, the DIRBE model predictions for these two bands were coadded and fitted with a power law emissivity model and extrapolated to the frequency coverage of FIRAS. The zodi model used here is among FIRAS data products on LAMBDA. 
For further details of the model derivation, see \citet{Fixsen02}.

\subsubsection{Emission lines}

The FIRAS detected 18 molecular and atomic lines emitted by interstellar gas. Since the FIRAS frequency resolution is much larger than the width of each of these lines, each line profile is effectively FIRAS's instrument response to a delta function. 
Among these 18 detected emission lines, not all of them have a discernible presence over the full sky. Most notable are the [C\,{\sevensize II}] and [N\,{\sevensize II}] lines, which exhibit a distinct gradient of intensity from the centre of the Galaxy to higher latitudes. Other emission lines, though detected, are weak in most of the sky except at the Inner Galaxy. By Inner Galaxy we mean the inner Galactic disk about half the distance to the edge of the Galaxy. 

Since the derivation of FIRAS line intensity maps on LAMBDA used their Galactic dust spectra, the FIRAS line intensity maps cannot be used here to remove emission line contribution. Instead, intensities of [C\,{\sevensize II}] and [N\,{\sevensize II}] emission were fit as parts of the overall model.

\subsubsection{Cosmic infrared background}\label{sec:cib_removal}

We removed the isotropic CIB signal from sky spectra by using results from three studies: For DIRBE measurements at 140 and 240~$\mu$m, we used estimates given by \citet{Hauser98} at 15 and 13~nW~m$^{-2}$~sr$^{-1}$. For the DIRBE band at 100~$\mu$m, we use the estimate given in \citet{Finkbeiner00} at 25~nW~m$^{-2}$~sr$^{-1}$, see also \citet{Lagache00}.
To remove the CIB from FIRAS sky spectra, we used the CIB model in \citet{Fixsen98}.

\subsection{Beam difference}\label{sec:beam_difference}

The three instruments that produced the data used in this study had different beam patterns. For example, the FIRAS used a quasi-optical multimode horn antenna to collect radiation from a 7$^\circ$ field of view \citep{Mather86}. The horn was designed in a trumpet bell shape to reduce response to off-axis radiation. As a result, when the beam profile was measured on the ground and in flight, it was found to have very low sidelobes over the two decades of frequency measured by FIRAS. 
The central portion of the beam ($\theta < 3\fdg $5) was approximated by a top hat since any slight azimuthal asymmetry should have been smoothed out by the rotation of the instrument along its own axis \citep{Mather93} during its operation. 

On the other hand, DIRBE was built with a goal to reject stray light to measure the absolute spectrum and angular distribution of the CIB. This goal was met by using a series of optical elements and baffle protections, among which the last field stop set the $0\fdg 7 \times 0\fdg 7$ instantaneous field of view for all spectral bands \citep{Silverberg93,DIRBE98}. To construct the ZSMA maps, the DIRBE Team calculated the zodiacal light intensity using the IPD model by \citet{Kelsall98} and subtracted it off from each weekly measurement. The remaining signal was averaged over time. In this way, the ZSMA maps preserved the original $0\fdg 7 \times 0\fdg 7$ angular resolution of the sky observation.

The dust map from the {\sl WMAP} was one of the products derived from Markov chain Monte Carlo fitting of temperature and polarization data \citep{Gold09}. Since their analysis used the band-averaged maps that were smoothed by a 1$^\circ$ Gaussian beam, the dust map had the same angular resolution. 

Since the FIRAS beam had the lowest common angular resolution achievable among all three data sets, we convolved the higher resolution DIRBE and {\sl WMAP} maps with the FIRAS beam to make them all 7$^\circ$ maps. Additionally, since \citet{Fixsen97b} found that the FIRAS beam was elongated in the scan direction by 
2$\fdg$4, we matched that pattern in the degraded DIRBE and {\sl WMAP} maps by convolving those data with an effective FIRAS beam.

\subsection{Map projection and spatial resolution}\label{sec:data_prep_hpx2csc}

Both DIRBE and FIRAS maps are organized in {\sl COBE} quadrilateralized spherical cube format (quad-cube, \citealt*{Chan75}, \citealt*{ONeill76}, \citealt*{White92}, and \citealt*{Calabretta02}). While the DIRBE maps are in quad-cube resolution level 9 (res9, 19$\fm$43 per pixel), the FIRAS maps are in quad-cube resolution level 6 (res6, 2$\fdg$59 per pixel). Different from the FIRAS or the DIRBE maps, the {\sl WMAP} dust map is in HEALPix (\citealt*{Gorski99}, \citealt*{Gorski05}, and \citealt*{Calabretta07}) resolution level 6 (res6, 54$\fm$97 per pixel). One way to reconcile these different formats and spatial resolutions is to carry out analysis in {\sl COBE} quad-cube res6. With this decision we hope to retain maximum amount of information contained in the original data sets and to achieve the highest common resolution possible. 

We re-binned DIRBE maps to res6. The {\sl WMAP} dust map was first converted into a quad-cube res9 map and then re-binned to res6. During {\sl WMAP}'s dust map conversion, we checked to make sure that no excessive artificial noise was introduced to the final map: By comparing the original HEALPix-projection map with the re-binned quad-cube-projection map, we found that 98.6~per~cent of the 49,152-coordinate pairs sampled gave no difference between the quad-cube and the HEALPix values. When there was a difference, the maximum was 0.0059~mK, which amounted to a 0.11~per~cent noise increase for the original HEALPix map.

\subsection{Gradient correction}

Since the production of FIRAS dust maps involved coadding interferograms, the value reported for each pixel is generally defined for a location within the pixel other than its defined centre. This positional difference requires additional correction to prepare the quad-cube res6 DIRBE and WMAP maps. Details of this technique are described in \citet{Fixsen97b}. In summary, we fit a second-degree surface function to the intensity and location information of a pixel and its immediate neighbors in one of the converted maps. We then use this function to predict emission at the FIRAS mean position for that particular pixel. Overall, we measure a 5~per~cent rms correction to each of the DIRBE maps at 100, 140 and 240~$\mu$m and to the {\sl WMAP} dust map.

\subsection{Color correction}

In accordance with the {\sl IRAS} convention \citep{IRAS88}, DIRBE photometric measurements were reported in MJy~sr$^{-1}$ at nominal wavelengths, assuming the source spectrum to be $\bmath{\nu} \bmath{\cdot} \bmath{I_\nu} = \mathrm{constant}$. Since each DIRBE band has a much wider bandwidth than a FIRAS channel, spectral shape could have changed enough that at the nominal wavelength the real intensity was significantly different from the normalized intensity. As a result, we include color corrections in the overall model, i.e., model predictions are compared with DIRBE measurements using the relation $\bmath{I}_{\bmath{\nu},\rmn{\, model}} = \bmath{I}_{\bmath{\nu},\rmn{\, DIRBE}} / K$, where $K$ is the color correction factor defined as 
\begin{equation}
K = \frac{\int(\bmath{I_\nu}/I_{\nu_0})_\rmn{actual} \bmath{\cdot} \bmath{R_\nu} \, \rmn{d}\,\nu}{\int(\nu_0/\bmath{\nu})_\rmn{quoted} \bmath{\cdot} \bmath{R_\nu} \, \rmn{d}\,\nu} \,.
\end{equation}
In this equation, $\int(\bmath{I_\nu}/I_{\nu_0})_\rmn{actual}$ is the specific intensity of the sky normalized to the intensity at frequency $\nu_0$ and $\bmath{R_\nu}$ is DIRBE relative system response at frequency $\bmath{\nu}$. The values of $\bmath{R_\nu}$ are documented in \citet{DIRBE98} Section 5.5.

\subsection{DIRBE uncertainties}

The DIRBE photometric system was maintained to $\sim$~1~per~cent accuracy by monitoring the internal stimulator during 10 months of cryogenic operation and observing the bright stable celestial sources during normal sky scans. It was absolutely calibrated against Sirius, NGC7027, Uranus, and Jupiter. 

Among different types of uncertainties identified by the DIRBE Team, those relevant to this work are standard deviations of intensity maps, detector gain and offset uncertainties, and zodi model uncertainties \citep{Hauser98, Kelsall98, Arendt98}. For bands 8--10, respectively, we use detector gains estimates at 0.135, 0.106 and 0.116~nW~m$^{-2}$~sr$^{-1}$, detector offsets at 0.81, 5 and 2~nW~m$^{-2}$~sr$^{-1}$, and zodi model uncertainties at 6, 2.3 and 0.5~nW~m$^{-2}$~sr$^{-1}$ \citep{Arendt98}. The process of re-binning the high resolution DIRBE maps into FIRAS resolution affects the standard deviations of the original maps only. The final uncertainty is the quadrature sum of the individual noise components.

\subsection{Temperature-intensity conversion}\label{sec:temp_intensity}

Foreground maps of the {\sl WMAP} production are reported in antenna temperature, $T_A$, in mK. On the other hand, maps produced by the DIRBE and the FIRAS Teams are reported in spectral intensity, $I_\nu$, in MJy~sr$^{-1}$. In the following analysis, the {\sl WMAP} dust map is converted into flux density values following $I_\nu = 2 \,\,(\nu / c)^2\,\, k\, T_A$, where $\nu$ is the effective frequency (93.5~GHz) of the dust map \citep{Gold09, Jarosik03} and $k$ is Boltzmann's constant.

%% file: analysis4ps_dust6.tex
\subsection{Overview of model fitting strategy}

The thermal emission of a dust grain can be modeled by a modified blackbody function: 
\begin{displaymath}
\bmath{I}_\rmn{dust}(\bmath{\nu}) = \tau\ \bmath{\epsilon_\nu}\bmath{\cdot}\bmath{B_\nu}(T_{\mathrm{dust}}) \,,
\end{displaymath}
where $\bmath{B_\nu}(T_\mathrm{dust})$ is the blackbody spectrum at temperature $T_\mathrm{dust}$, $\bmath{\epsilon_\nu} = (\bmath{\nu}/\nu_0)^\beta$ is the emissivity with spectral index $\beta$, and $\tau$ is the optical depth normalized to frequency $\nu_0 = 900$~GHz.

In addition to measuring thermal dust emission, the prepared FIRAS spectra retain contributions from [C\,{\sevensize II}] and [N\,{\sevensize II}] emission due to the lack of precise all-sky templates. As a result, the two emission lines are modeled as: 
\begin{displaymath}
\bmath{I}_\rmn{[C\,_{II}]}(\bmath{\nu}) = \rmn{[C\,_{II}] \, _{intensity}} \,\, \bmath{f}_\rmn{[C\,_{II}]}(\bmath{\nu})\,,\\
\end{displaymath}
and
\begin{displaymath}
\bmath{I}_\rmn{[N\,_{II}]}(\bmath{\nu}) = \rmn{[N\,_{II}] \, _{intensity}} \,\, \bmath{f}_\rmn{[N\,_{II}]}(\bmath{\nu})\,,
\end{displaymath}
where $\bmath{f}(\bmath{\nu})$ is the synthetic line profile determined by FIRAS response to a delta-function signal. Together, the full model has the form 
\begin{equation} 
\bmath{I}_\mathrm{total}=\bmath{I}_\mathrm{dust}+\bmath{I}_\mathrm{[C\,_{II}]}+\bmath{I}_\mathrm{[N\,_{II}]} \,. 
\end{equation}

Each full model is fit to the data by minimizing a three-part $\chi^2$ with each part corresponding to one of the three data sets:
\begin{equation} \chi^2 = \chi^2_{\mathrm{DIRBE}} + \chi^2_{\mathrm{FIRAS}} + \chi^2_{\mathrm{WMAP}} \,, \end{equation}
where \begin{equation} \chi^2_{\mathrm{instrument}} = \sum\limits_{i,j} (\bmath{I}_{\mathrm{obs}}-\bmath{I}_{\mathrm{mdl}})_i \,\,  (\mathbfss{M}^{-1})_{ij} \,\, (\bmath{I}_{\mathrm{obs}}-\bmath{I}_{\mathrm{mdl}})_j \,, \end{equation}
Here, $\bmath{I}_{\mathrm{obs}}$ is the observed spectral intensity, $\bmath{I}_{\mathrm{mdl}}$ is the model prediction, and \mathbfss{M} is the covariance matrix of the respective data set. 

In the following sections, we fit one-component dust models to spectra of fixed (Section \ref{sec:analysis}) and different (Section \ref{sec:one_comp_var_res}) size sky regions. In the former case, spectra retain the 7$^\circ$ angular size of FIRAS pixels; in the latter case, the 7$^\circ$ spectra are averaged by various amounts to increase signal-to-noise of the final spectra. Chi-square per degree of freedom values, $\chi^2_\mathrm{dof}$, are used to assess the quality-of-fit of a model to each spectrum. In particular, the one-component fixed-$\beta$ models have $214-4 = 210$ degrees of freedom while the free-$\beta$ model has 209 degrees of freedom. Adopting a 10~per~cent probability cut-off, it corresponds to a $\chi^2_\mathrm{dof} \la 1.13$ for both fixed- and free-$\beta$ models.

\subsection{Fitting spectra of 7$^\circ$ sky regions}\label{sec:analysis}

We fit to the spectrum at each $7^\circ$ pixel one-component models with fixed $\beta$ in the range $1.4-2.6$ at 0.1 increment. The fits have acceptable values ($\chi^2_\mathrm{dof} \le 1.13$) over most of the sky except at the Galactic plane.

\subsubsection{Quality-of-fit of models}

\begin{figure}
	\centering
	\includegraphics[scale=0.55]{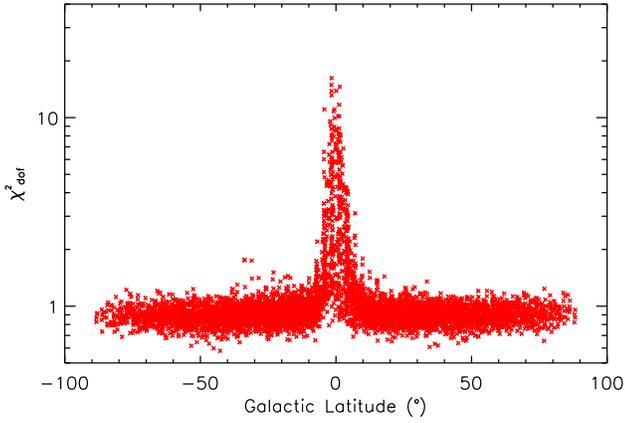}
	\caption{$\chi^2_{\mathrm{dof}}$ vs. Galactic latitude. The value of $\chi^2_{\mathrm{dof}}$ is obtained from fitting one-component $\beta = 2.0$ model to 7$^\circ$ spectra covering 98.68~per~cent area of the full sky where FIRAS data are available.}
	\label{fig:f1}
\end{figure}

\begin{figure}
	\centering
	\includegraphics[scale=0.55]{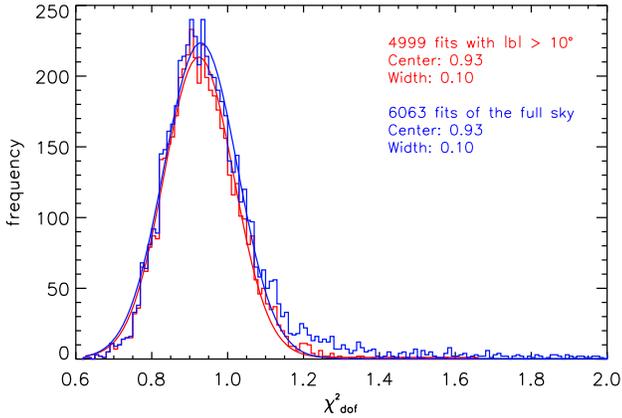}
	\caption{$\chi^2_{\mathrm{dof}}$ distributions of the 7$^\circ$ fits using one-component $\beta = 2.0$ model. The red distribution includes only pixels at Galactic latitudes $|b|>10^\circ$; the blue distribution includes all 6063 pixels at all Galactic latitudes. Best-fitting parameters of the Gaussians are printed in respective colors.} 
	\label{fig:chi2_alpha2_hist}
\end{figure}

As an example, Fig. \ref{fig:f1} presents $\chi^2_{\mathrm{dof}}$ as a function of Galactic latitude for the $\beta = 2.0$ model. It shows that most fits at $|b| \ga 10^\circ$ have $\chi^2_{\mathrm{dof}} \approx 1$ and that fits at $|b| \la 10^\circ$ have a $\chi^2_{\mathrm{dof}} > 1$. Fig. \ref{fig:chi2_alpha2_hist} compares the distribution of $\chi^2_{\mathrm{dof}}$ at $|b| \ga 10^\circ$ with the distribution of $\chi^2_{\mathrm{dof}}$ for the entire sky. 
That both distributions are well approximated by a Gaussian indicates that the fits don't have a significant systematic bias. The widths of the distributions are as expected (~0.1) for a distribution of random data with 210 degrees of freedom. That the distributions center at 0.93 means that statistical errors of the data are slightly overestimated by $\sim 7\%$, and that the $\chi^2_\mathrm{dof}$ cutoff is really at $\approx 1.21$ with a probability of $< 10\%$. Because the uncertainties of the FIRAS data include some systematic effects, we do not feel at liberty to reduce the uncertainty. Based on the $\chi^2_{\mathrm{dof}} \le 1.13$ cut, the model is a good fit to the data for over 87~per~cent of the full sky area and is rejected by the data at the Galactic plane. 

Readers interested in the best-fitting parameters ($T_\mathrm{dust}$, $\tau$, [C\,{\sevensize II}] and [N\,{\sevensize II}] intensities), their uncertainties and correlations for each of the aforementioned $\beta$ models are referred to \citet{Liang11}. 

\subsubsection{Dependence of best-fitting parameters on the signal-to-noise of data}

\begin{figure}
	\centering
    \includegraphics[scale=0.41]{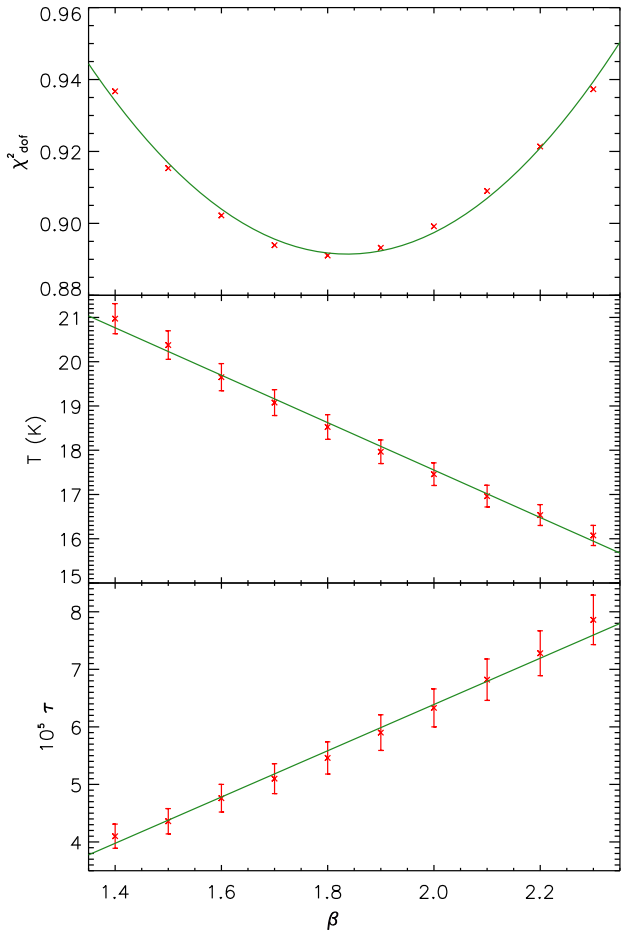}
  \caption{$\chi^2_\mathrm{dof}$, $T_\mathrm{dust}$ and $\tau$ as a function of $\beta$. Each data point on the $T_\mathrm{dust}$ and $\tau$ plots is the best-fitting value of the corresponding one-component fixed-$\beta$ model to the 7$^\circ$ spectrum measured in the direction $l = 63\fdg 78 \,\,\mathrm{and} \,\, b = -11\fdg 53$. This set of plots serves as an example of high signal-to-noise fits. The green curve fits the best-fitting values as a function of $\beta$.}
 	\label{fig:discrete_alpha_pixel_fit_pix0001}
\end{figure}

\begin{figure}
	\centering
    \includegraphics[scale=0.41]{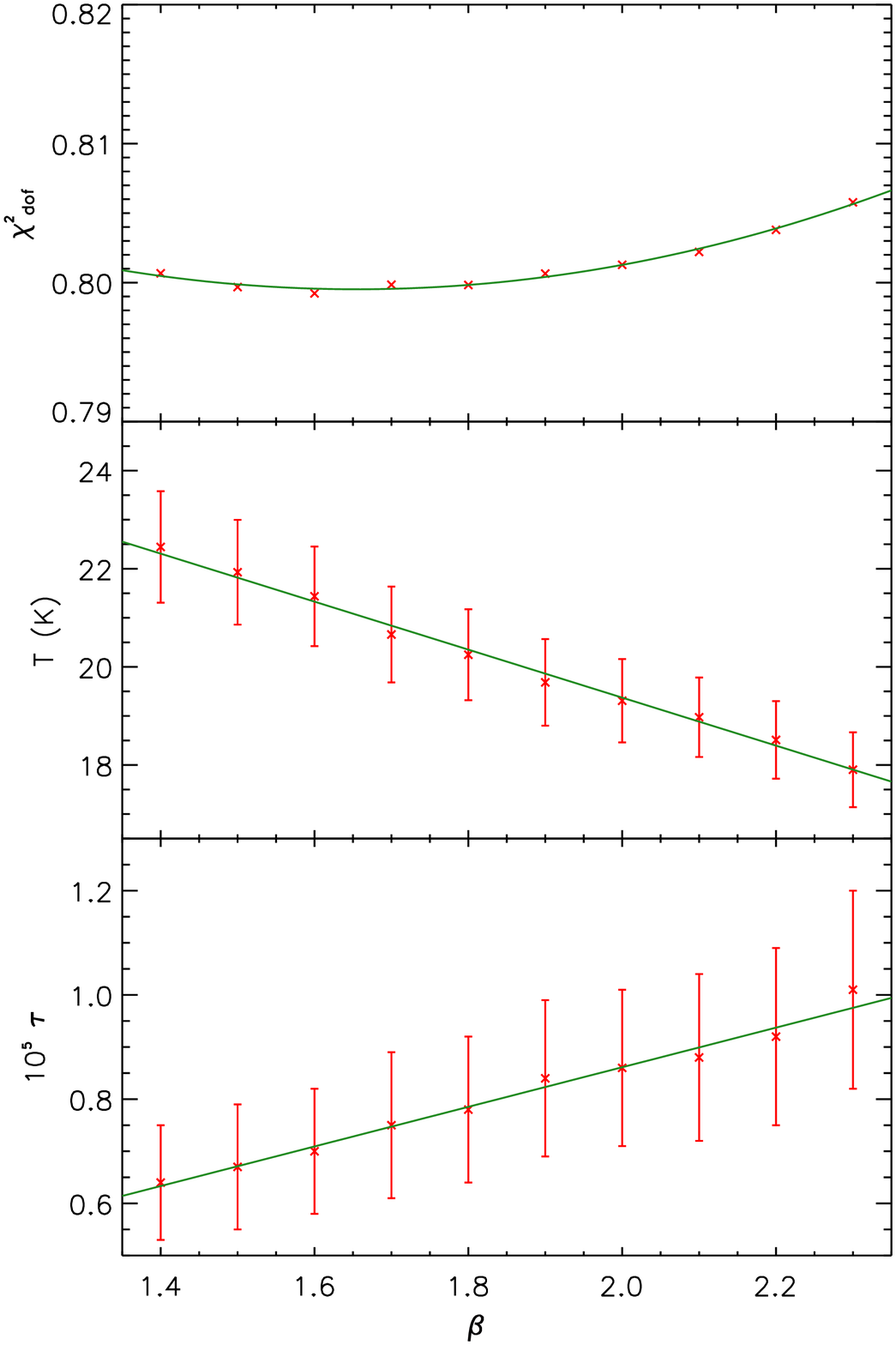}
  \caption{$\chi^2_\mathrm{dof}$ , $T_\mathrm{dust}$ and $\tau$ as a function of $\beta$. Each data point on the $T_\mathrm{dust}$ and $\tau$ plots is the best-fitting value of the corresponding one-component fixed-$\beta$ model to the 7$^\circ$ spectrum measured in the direction $l = 254\fdg 32 \,\,\mathrm{and} \,\, b = 65\fdg 08$. This set of plots serves as an example of low signal-to-noise fits. The green curve fits the best-fitting values as a function of $\beta$.}
 	\label{fig:discrete_alpha_pixel_fit_pix3672}
\end{figure}

Figs. \ref{fig:discrete_alpha_pixel_fit_pix0001} and \ref{fig:discrete_alpha_pixel_fit_pix3672} present $\chi^2_{\mathrm{dof}}$, $T_\mathrm{dust}$ and $\tau$ of the best-fitting $\beta$ models for two spectra: one has high signal-to-noise (HSN) and is at a low Galactic latitude;  the other one has low signal-to-noise (LSN) and is at a high Galactic latitude. For both cases, plots of $\chi^2_{\mathrm{dof}}$ vs. $\beta$  show that models with a wide range of different $\beta$ values can fit the data well. In the HSN case, a curve fit to $\chi^2_{\mathrm{dof}}$ as a function of $\beta$ is a concave up parabola, with the minimum $\chi^2_{\mathrm{dof}} = 0.89$ at $\beta \approx 1.8$. The difference between $\chi^2_{\mathrm{dof}}$ at $\beta = 1.8$ and that at $\beta = 2.0$ is $\Delta \chi^2_\mathrm{dof} = 0.01$. At 210 degrees of freedom, this means a $\Delta \chi^2$ of $\sim 2.1$, a 2-sigma difference. 
In the LSN case, the best-fitting curve to $\chi^2_{\mathrm{dof}}$ vs. $\beta$ is a much flatter parabola over $1.4 \le \beta \le 2.3$ with the minimum $\chi^2_\mathrm{dof} = 0.80$ at $\beta = 1.6$. The difference between $\chi^2_{\mathrm{dof}}$ at $\beta = 1.8$ and $\beta = 2.0$ is $\sim 0.002$.

Although models with different $\beta$ have only a small difference in $\chi^2_\mathrm{dof}$, the best-fitting $T_\mathrm{dust}$ and $\tau$ are different significantly in the HSN case: At $\beta = 2.0$, the best-fitting dust temperature is $T_\mathrm{dust} = 17.5 \pm 0.26$~K, compared to $T_\mathrm{dust} = 18.5 \pm 0.28$~K at $\beta = 1.8$. This difference in temperature is larger than the sum of their errors. Similarly, the difference in $\tau$ of the two $\beta$ models is larger than the sum of their errors. On the contrary, in the LSN case, the difference in the best-fitting $T_\mathrm{dust}$ and $\tau$ of $\beta = 1.8$ and $\beta = 2.0$ models are well within the uncertainties of the respective parameters.

\begin{figure}
	\centering
    \includegraphics[scale=0.80]{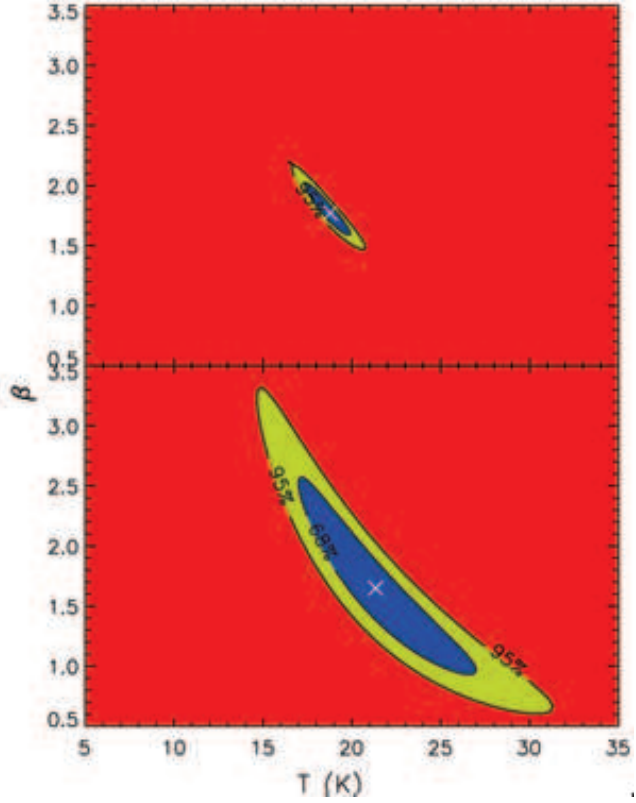}
  \caption{68 and 95~per~cent probability contours in the $T_\mathrm{dust}$-$\beta$ space for a high signal-to-noise spectrum (upper plot, measured in the direction $l = 63\fdg 78 \,\,\mathrm{and} \,\, b = -11\fdg 53$) and a low signal-to-noise spectrum (lower plot, $l = 254\fdg 32 \,\,\mathrm{and} \,\, b = 65\fdg 08$). In both plots, the white cross represents location of the minimum $\chi^2$; the blue area is the 68~per~cent confidence region; and the green area encloses the 95~per~cent confidence region. These plots demonstrate the effect of measurement noise on the degeneracy between $\beta$ and $T_\mathrm{dust}$ in spectral model fitting. For the high signal-to-noise spectrum (upper plot), the 68~per~cent confidence region is at $1.7<\beta<1.9$ and $17.9\mathrm{\,\,K \,\,}<T_\mathrm{dust}<19.5$~K; for the low signal-to-noise spectrum (lower plot), the 68~per~cent confidence region has a much wider extent, at $1.2<\beta<2.2$ and $18.5\mathrm{\,\,K \,\,} < T_\mathrm{dust}<24.7$~K.}
 	\label{fig:confidence_regions_pix0001_and_pix3672}
\end{figure}

These results demonstrate the sensitivity of the fits to measurement errors. The existence of measurement noise inevitably causes a high degree of degeneracy between emissivity spectral index and dust temperature in the fits. While it is difficult to break this degeneracy, high signal-to-noise data help. Fitting data with high signal-to-noise results in well constrained parameters, which means that the choice of an $\beta$ model can cause statistically significant differences in the predictions of these parameters. On the other hand, fitting low signal-to-noise spectra results in small difference in $\chi^2_\mathrm{dof}$ and large error bars of the best-fitting parameters. The results cannot be used to differentiate models with different fixed values of $\beta$. This is demonstrated in Fig. \ref{fig:confidence_regions_pix0001_and_pix3672}, which shows that the 68 and 95~per~cent confidence contours of a HSN fit enclose much smaller regions in the $T$-$\beta$ space than those of a LSN fit.

\begin{figure}
	\centering
	\begin{tabular}{c}
    \includegraphics[scale=0.31]{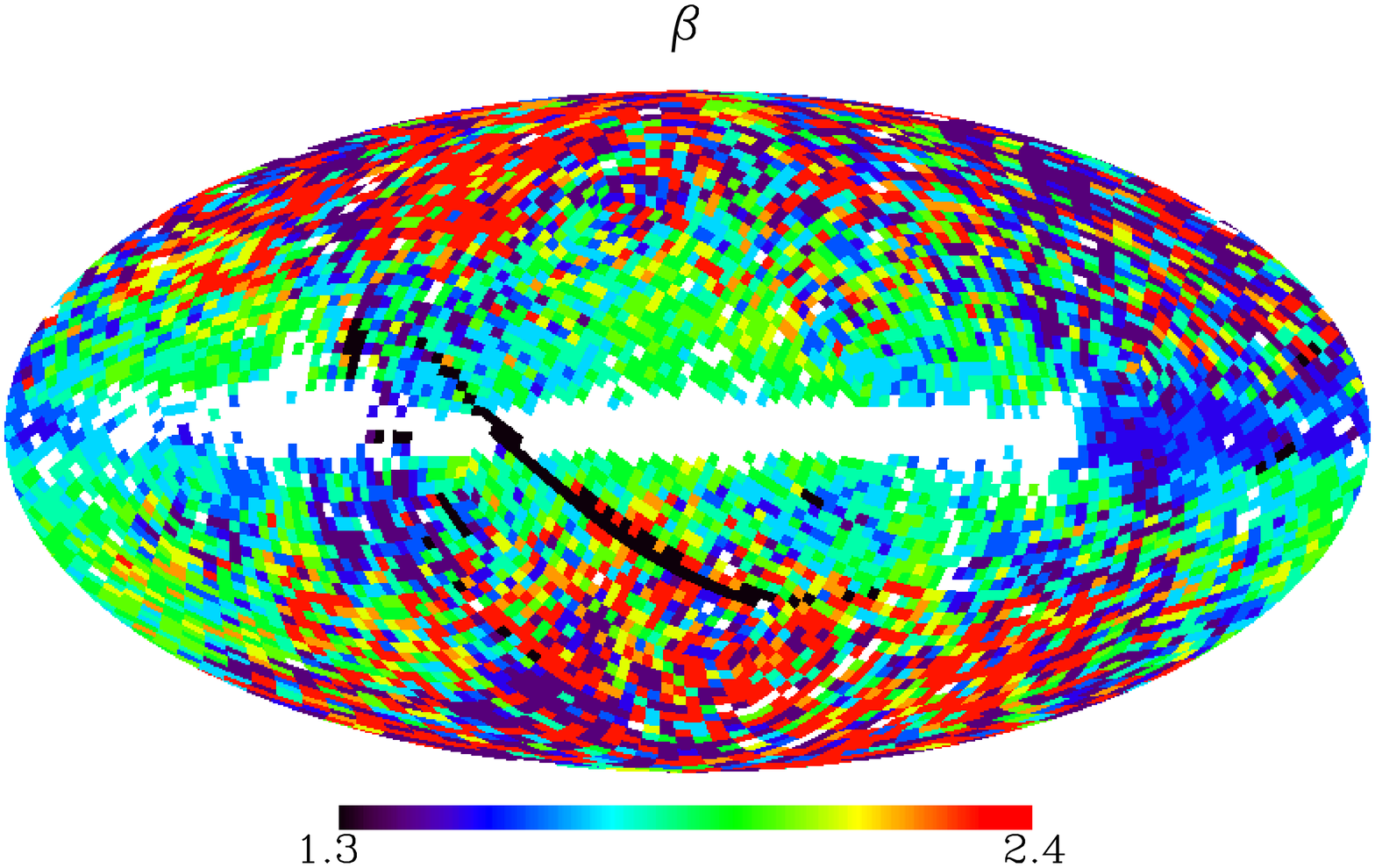}\\	
    \hspace{2pc}\\
		\includegraphics[scale=0.31]{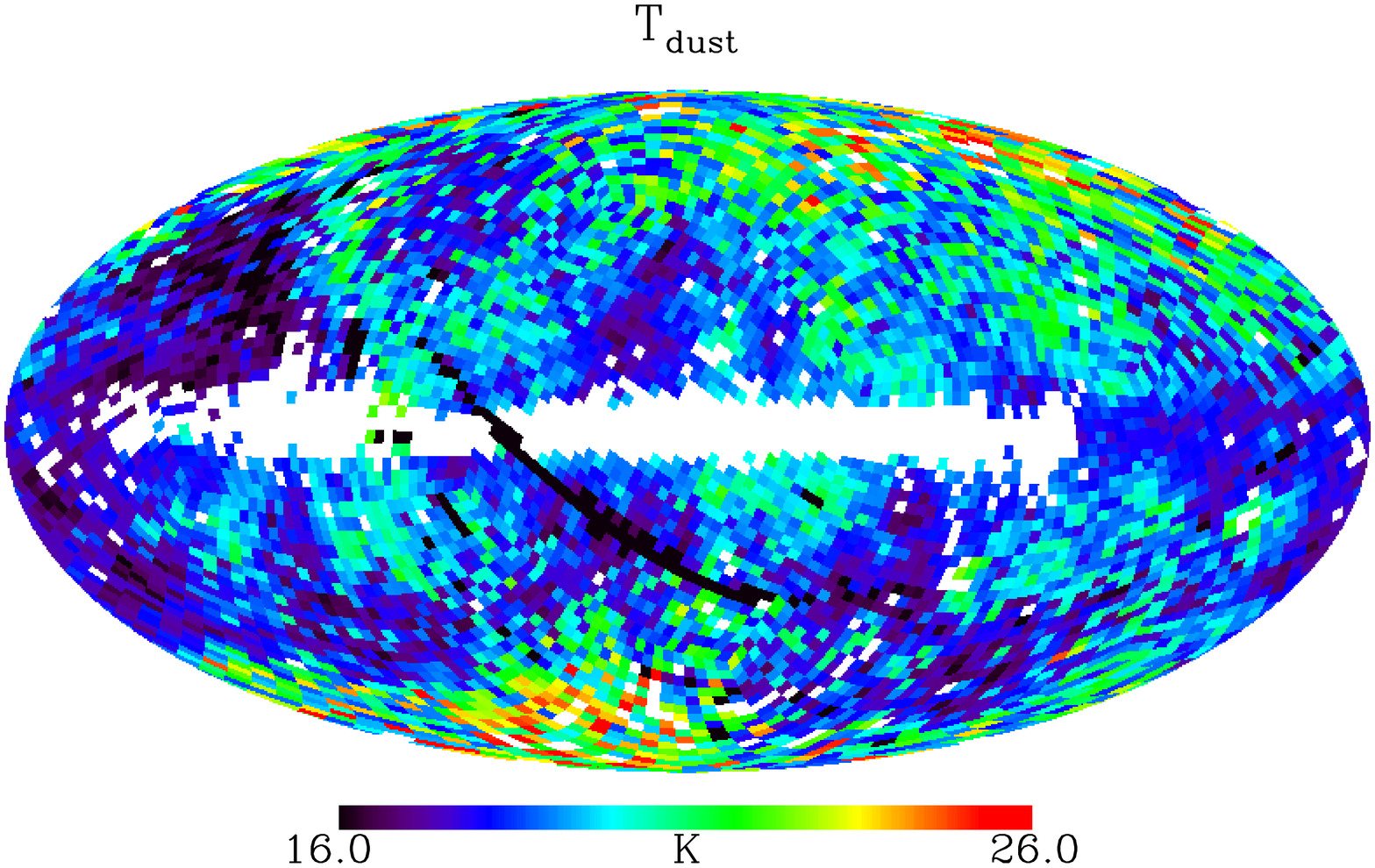}\\
	\end{tabular}
  \caption{Sky maps of $\beta$ and $T_\mathrm{dust}$. We fit all 6000 spectra with one-component fixed-$\beta$ models, where $1.4 \le \beta \le 2.3$ at 0.1 increment, identify the minimum-$\chi^2$ model for each spectrum, and use those models to construct these maps. In Galactic coordinates Mollweide projection, the centre of each map is the Galactic centre. The upper and lower ends of the minor axis are $+90^\circ$ and $-90^\circ$ latitudes respectively, and the left and right ends of the major axis represent $+180^\circ$ and $-180^\circ$ longitudes respectively. Pixels that correspond to fits with a $\chi^2$ less than 10~per~cent probability are masked in white. The group of black pixels that slant from the centre of the upper left quadrant (North Ecliptic Pole, NEP) to the centre of the lower right quadrant (South Ecliptic Pole, SEP) are positions where FIRAS did not provide data.}
 	\label{fig:discrete_alpha_pixel_fit_skymaps}
\end{figure}

Figs. \ref{fig:discrete_alpha_pixel_fit_skymaps} shows sky maps of $\beta$ and $T_\mathrm{dust}$ that correspond to the minimum-$\chi^2_{\mathrm{dof}}$ model among all models tested at each pixel. 
Both maps show greater consistency in value at low latitudes and more fluctuations around the Galactic poles. That consistent values appear in the region surrounding the Inner Galaxy is reasonable because star formation as the most important heat source for dust takes place in the Galactic disk and at the bulge. That large fluctuations appear near the poles, on the other hand, has to do with lower signal-to-noise data in these regions compared to those measured at lower latitudes. This happens because few dust grains exist at high latitudes, and they do not emit as strongly as those close to the Galactic disk. 

Since low signal-to-noise data cannot give adequate constraint to model parameters and exacerbates the degeneracy between dust temperature and spectral index, in order to construct the best dust model, we need to find ways to increase the signal-to-noise of the data. 

%% file: one_comp_var_res_dust6.tex
\subsection{Fitting averaged spectra of different-size sky regions}\label{sec:one_comp_var_res}

Taking average of the high-latitude spectra based on latitudinal or longitudinal divisions of the sky can tighten the constraint on model parameters. However, such divisions are based on our expectations of the distribution of Galactic dust. They are not optimal because our knowledge of the dust distribution is incomplete. In our experiments \citep{Liang11}, $\chi^2$ of regional fits are much higher than $\chi^2$ of fits to the individual $7^\circ$ spectra that comprise the regional averages. Since larger sky regions include different types of dust emission spectra, the steep increase in $\chi^2$ value means that the averaging has achieved a sufficient signal-to-noise ratio and that spectral variation has become statistically important. In order to preserve information on spectral variation in the model, the amount of spatial averaging needs to reflect the signal-to-noise of the data. 

One way to do this is to base the amount of spectral averaging on signal-to-noise of the averaged spectrum. Starting with the base level, where a pixel's own spectrum is used to fit a model, if the fit does not give well constrained parameters due to inadequate signal-to-noise, the procedure goes on to fit the average of the original spectrum and its eight immediate neighbors. This process of involving more of the adjacent spectra to form a new average goes on until the derived parameters are sufficiently constrained. In this way, results from fits done at the base level have a spatial resolution of 6.71~$\sq^2$. At the next level, results have a spatial resolution of 60.37~$\sq^2$, and so on.

\subsubsection{$T_\mathrm{dust}/\delta T_\mathrm{dust}$ constraint on fixed-$\beta$ models}\label{sec:T_SN_constraint}

\begin{figure}
	\centering
	\begin{tabular}{c}
   	  \includegraphics[scale=0.54]{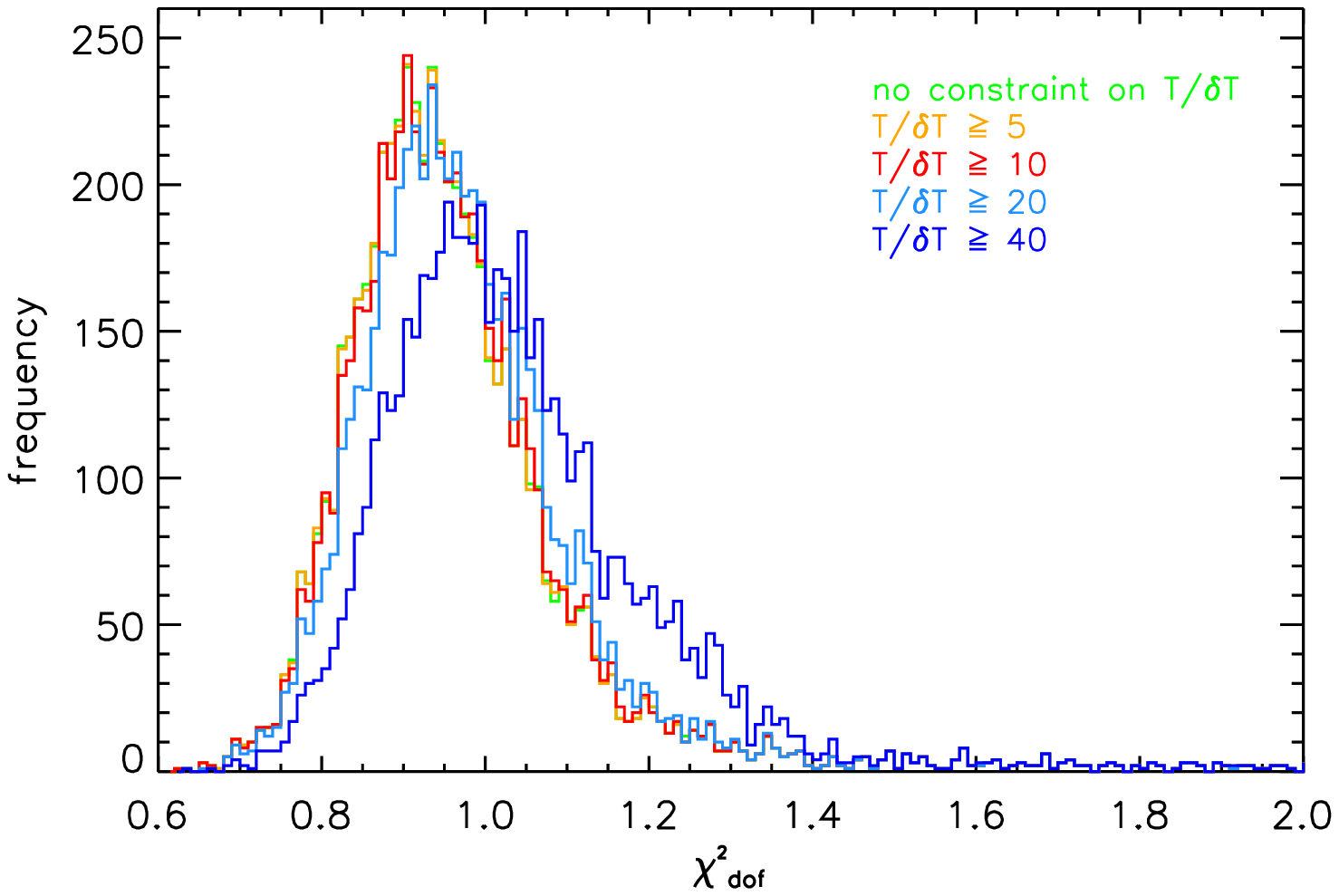}\\
	  \hspace{5pc}\\
	  \includegraphics[scale=0.54]{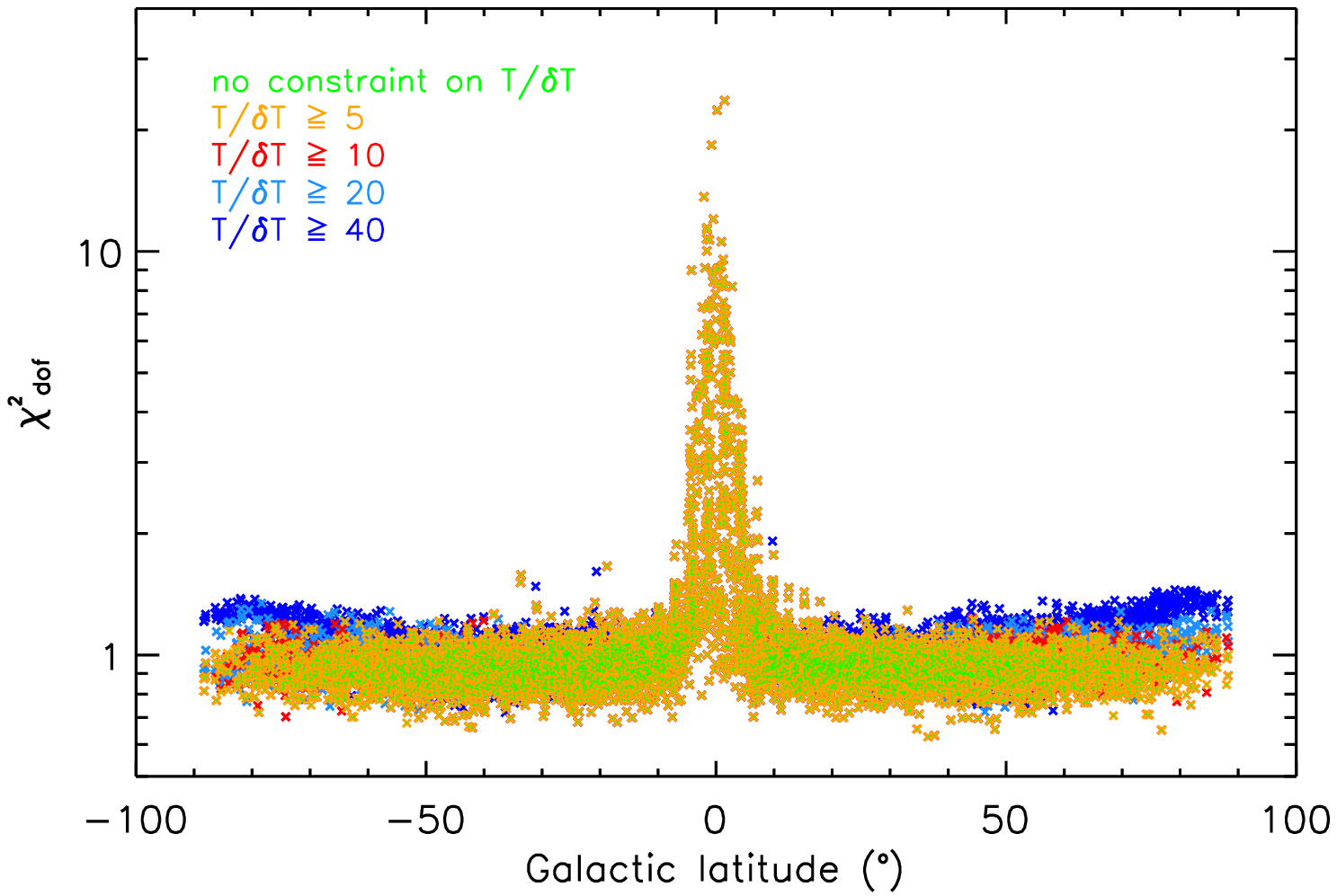}\\
	\end{tabular}
	\caption{$\chi^2_\mathrm{dof}$ distributions (upper) and $\chi^2_\mathrm{dof}$ vs. Galactic latitude (lower) of the all-sky one-component $\beta = 2.0$ fits that satisfy $T_\mathrm{dust}/\delta T_\mathrm{dust} \ge$ 5, 10, 20 and 40, respectively. With more restrictive $T_\mathrm{dust}/\delta T_\mathrm{dust}$ requirements, $\chi^2_\mathrm{dof}$ of high-latitude ($|b|>60^\circ$) fits migrate from the range $0.7-1.0$ to the range $0.8-1.3$. }
	\label{fig:chi2_vs_b_and_dist}
\end{figure}

All-sky one-component fixed-$\beta$ models with $\beta$ in the range $1.4-2.6$ and lower limit of $T_\mathrm{dust}/\delta T_\mathrm{dust}$ at 5, 10, 20 and 40 are obtained separately. In general, one-component fixed-$\beta$ models with different lower limits on the $T_\mathrm{dust}/\delta T_\mathrm{dust}$ values can fit most spectra except those at the Galactic plane. A more restricted lower limit on the $T_\mathrm{dust}/\delta T_\mathrm{dust}$ values requires a greater amount of spatial averaging. 
As dust spectra with refined differences in shape are binned together and the uncertainties of the averaged spectrum are reduced, we subject the comparison between the averaged spectrum and the modified blackbody function to an increasingly restricted standard. The result is a steep increase in $\chi^2$ value. 
The lower plot of Fig. \ref{fig:chi2_vs_b_and_dist} demonstrates this relation by plotting $\chi^2_\mathrm{dof}$ as a function of Galactic latitude for the case of $\beta = 2.0$. Note that $\chi^2_\mathrm{dof}$ of high-latitude ($|b|>60^\circ$) fits move from $0.7-1.0$ to $0.8-1.3$ as lower limit on $T_\mathrm{dust}/\delta T_\mathrm{dust}$ starts with none and increases to 40.

Histograms of the all-sky collections of $\chi^2_\mathrm{dof}$ for different limits on $T_\mathrm{dust}/\delta T_\mathrm{dust}$ are presented in the upper plot of Fig. \ref{fig:chi2_vs_b_and_dist}. For the case of $T_\mathrm{dust}/\delta T_\mathrm{dust} \ge$ 5, 10 and 20, the three histograms have the shape of a Gaussian, peak at 0.93, 0.94 and 0.96 respectively, and all have a width of 0.10. At $T_\mathrm{dust}/\delta T_\mathrm{dust} \ge$ 40 the histogram peaks at 1.00, has a width of 0.12, and a thick tail in the range $1.2 < \chi^2_\mathrm{dof} < 1.4$. This 
means that at the level of $T_\mathrm{dust}/\delta T_\mathrm{dust} \ge$ 40, it is no longer adequate to use a modified blackbody function to describe variations in the dust spectral shape.

Imposing a more stringent limit on $T_\mathrm{dust}/\delta T_\mathrm{dust}$ leads to a variety of spatial resolutions in each all-sky collection of fits. A more stringent requirement on $T_\mathrm{dust}/\delta T_\mathrm{dust}$ means lower spatial resolutions for fits at high latitudes. For the one-component $\beta = 2.0$ model with  $T_\mathrm{dust}/\delta T_\mathrm{dust} \ge 5$, only 0.59~per~cent of the total sky area require fits with a 60.37~$\sq^2$ resolution instead of the default 6.71~$\sq^2$. For a 10~per~cent constraint on $T_\mathrm{dust}$, 0.47~per~cent area of the full sky require fits to be at 167.70~$\sq^2$ resolution, 7.24~per~cent at 60.37~$\sq^2$ resolution, and the rest at 6.71~$\sq^2$ resolution. A balance between having adequate constraint on parameters, preserving as many valid models as possible, and keeping regional sizes low can be achieved at the $T_\mathrm{dust}/\delta T_\mathrm{dust} \ge$ 10 level.

\begin{figure}
	\centering
	\begin{tabular}{c}
    	  \includegraphics[scale=0.54]{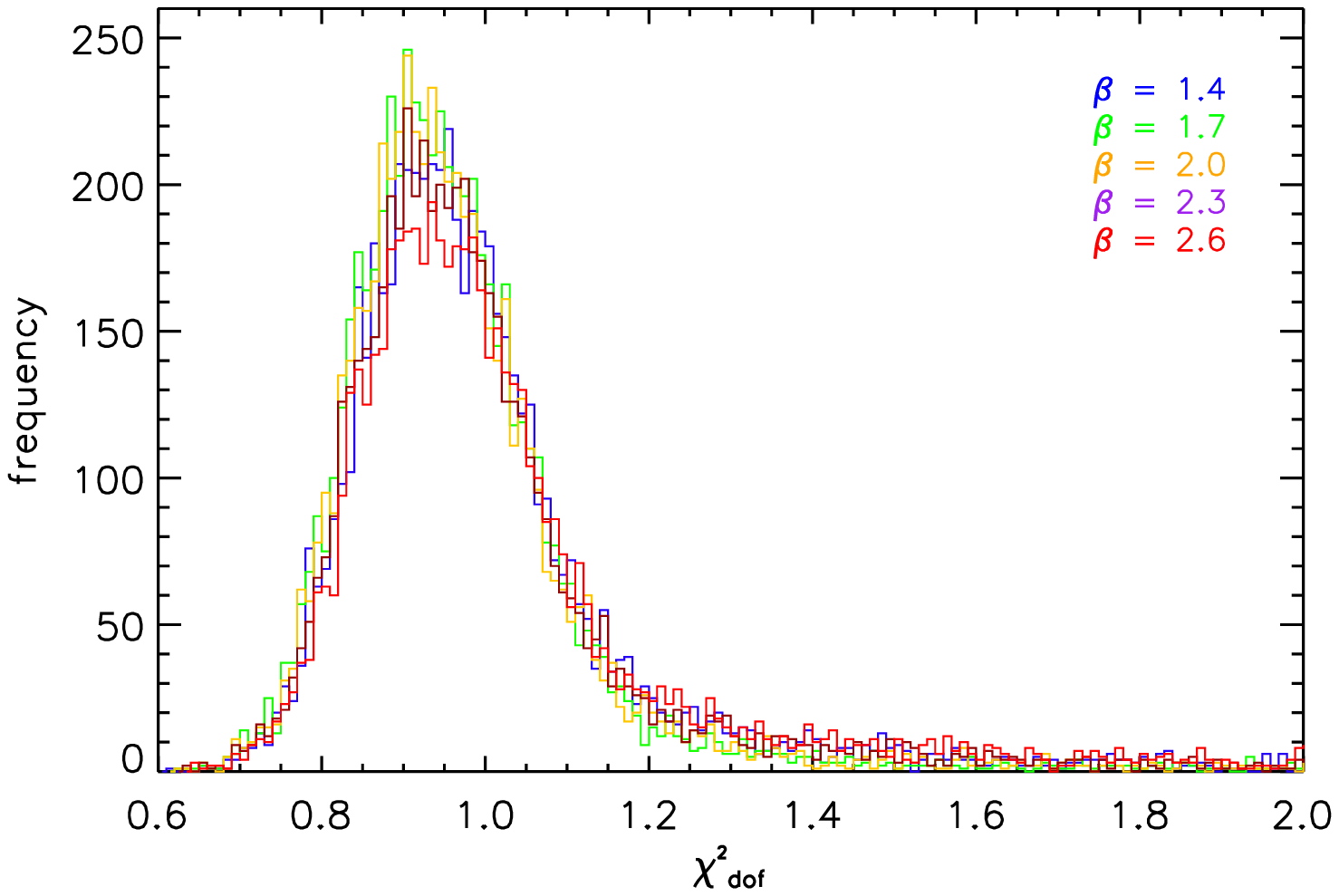}\\
	  \hspace{2pc}\\
	  \includegraphics[scale=0.54]{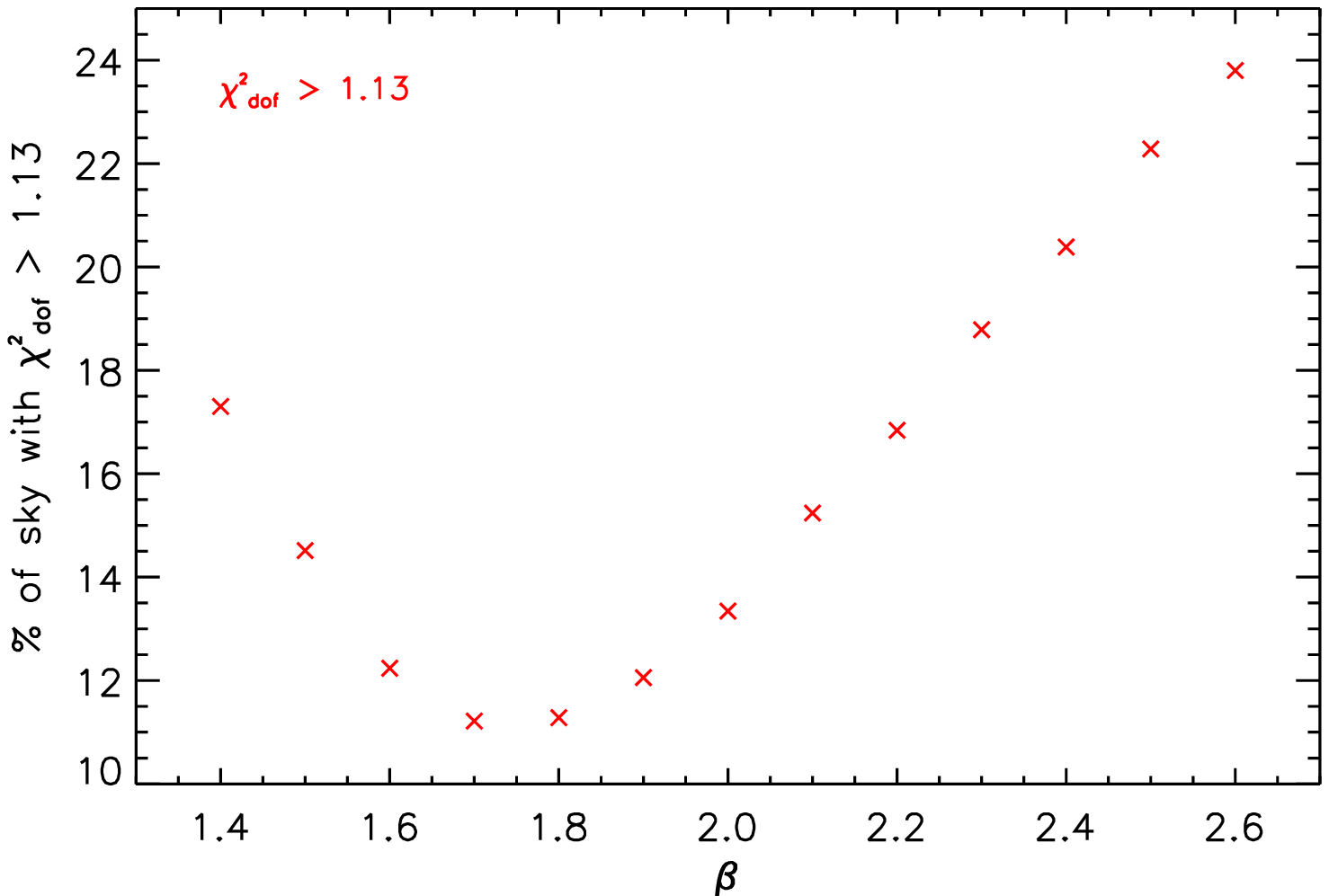}\\
	\end{tabular}
	\caption{Upper: $\chi^2_\mathrm{dof}$ distributions of the best-fitting one-component fixed-$\beta$ models with $\beta$ in the range 1.4 -- 2.6 at 0.3 increment and satisfy $T_\mathrm{dust}/\delta T_\mathrm{dust} \ge$ 10. Lower: Percentage area of the full sky that cannot be fit by a one-component model with fixed emissivity spectral index ($\chi^2_\mathrm{dof}$ cutoff corresponds to 10~per~cent probability). This plot shows that $\beta = 1.7$ models can fit the largest amount of data (87.6~per~cent area of the full sky).}
	\label{fig:chi2_hist_all_alphas}
\end{figure}

\begin{figure*}
	\centering
  {\bf One-component $\bmath{\beta = 1.7}$ fits with $T_\mathrm{dust}/\delta T_\mathrm{dust} \ge 10$}	

	\begin{tabular}{cc}
		\hspace{1pc}\\

    \includegraphics[scale=0.31]{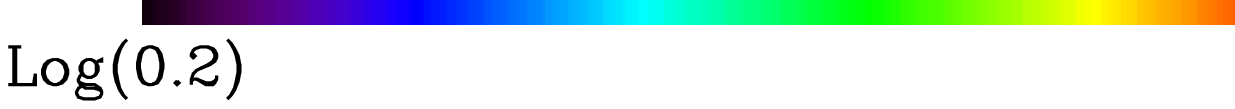}& 
    \includegraphics[scale=0.31]{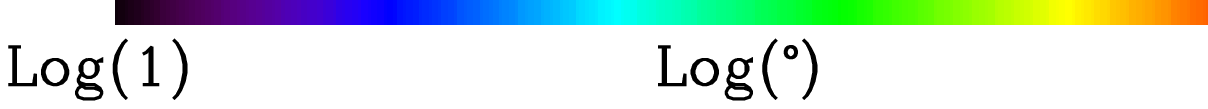}\\
	\hspace{1pc}\\
    \includegraphics[scale=0.31]{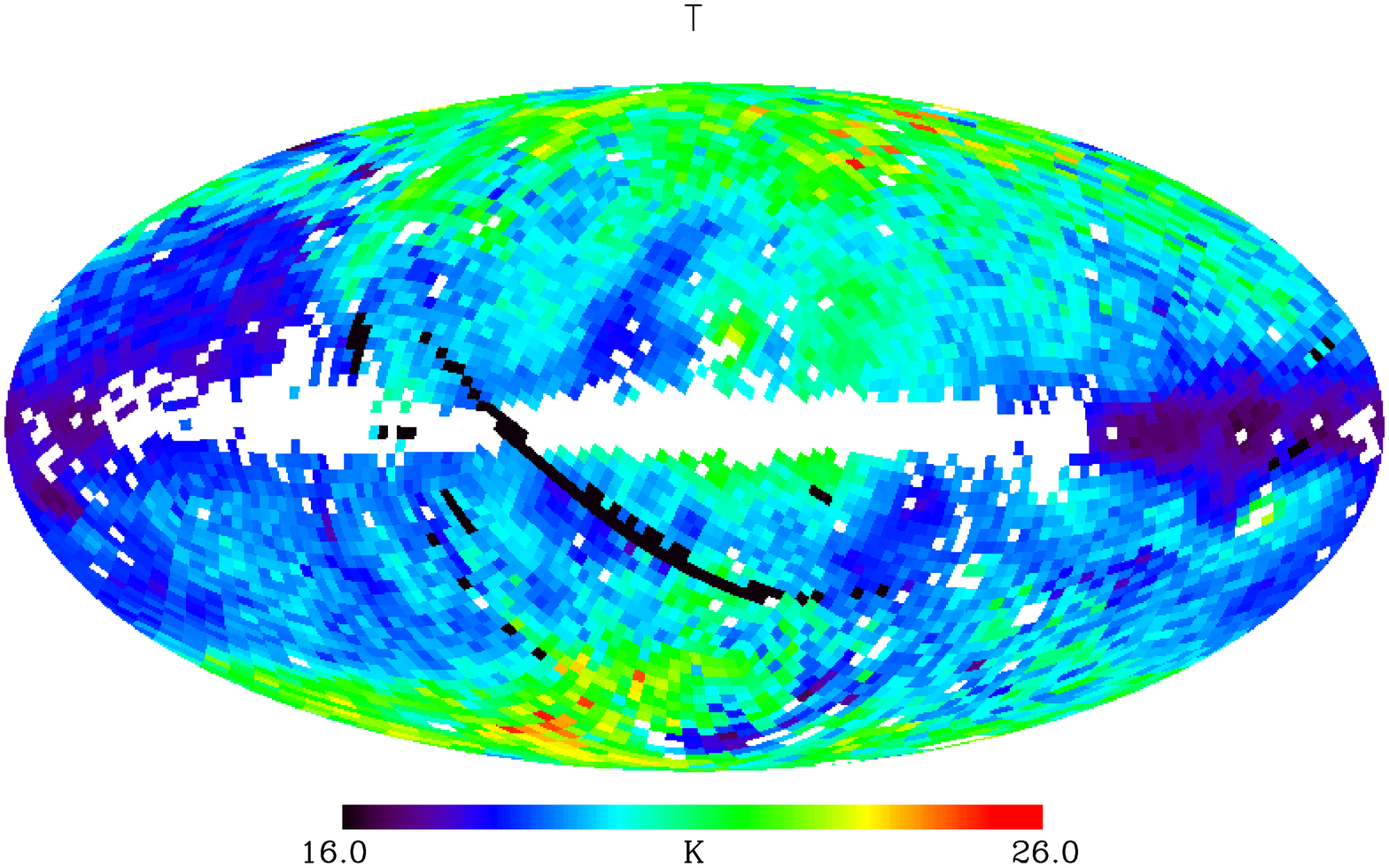}&
    \includegraphics[scale=0.31]{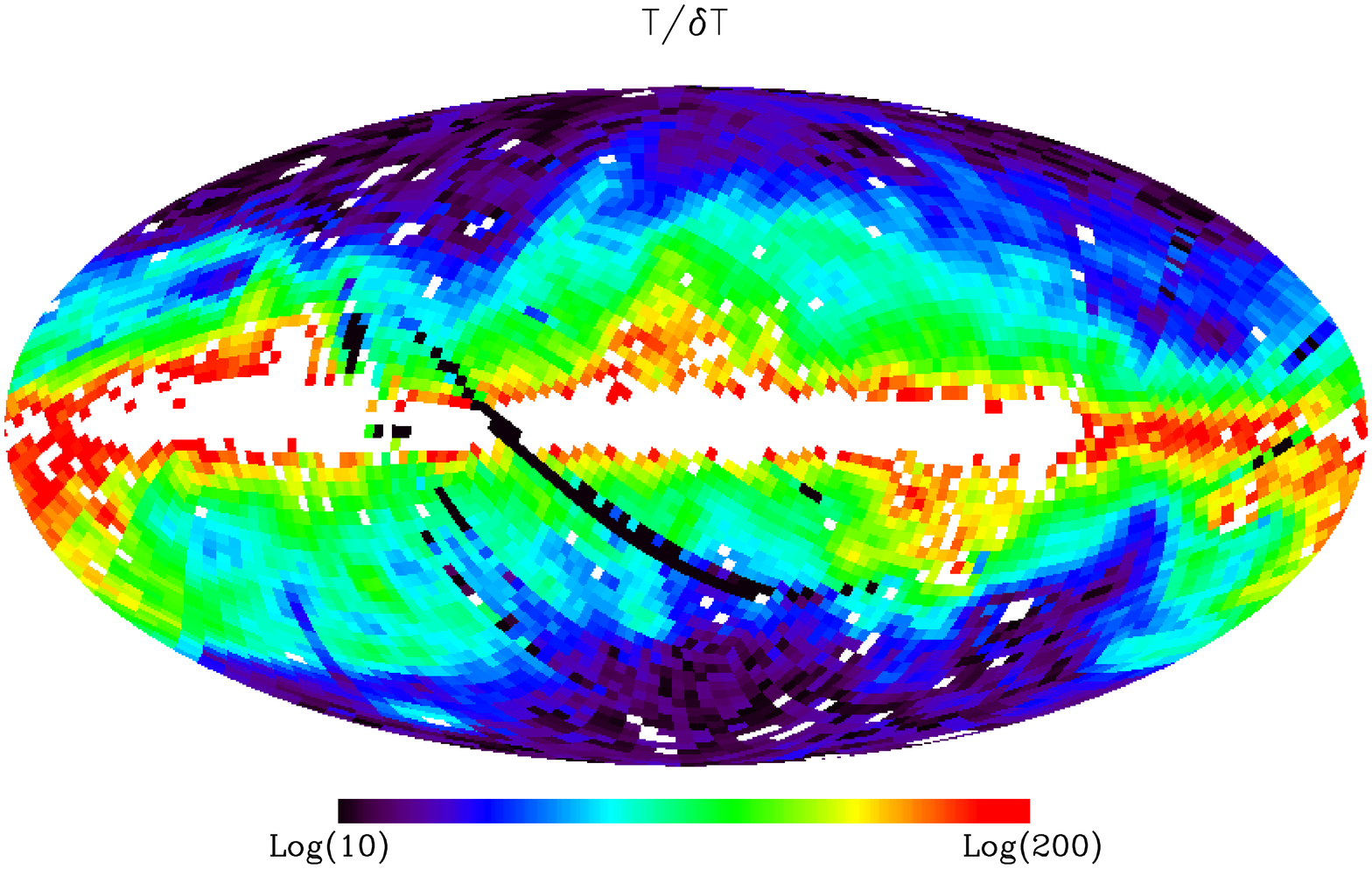}\\
	\hspace{1pc}\\
    \includegraphics[scale=0.31]{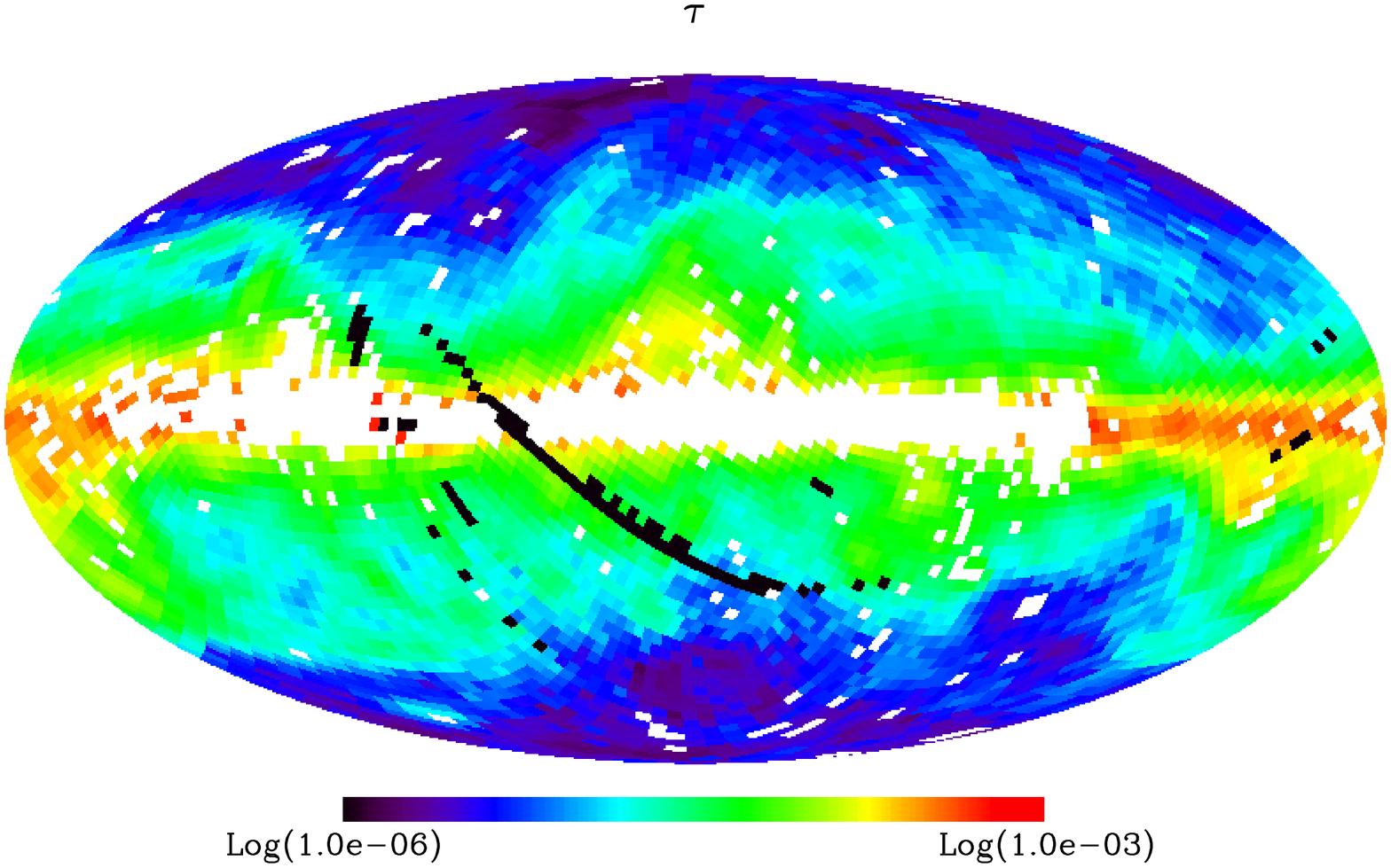}&	
    \includegraphics[scale=0.31]{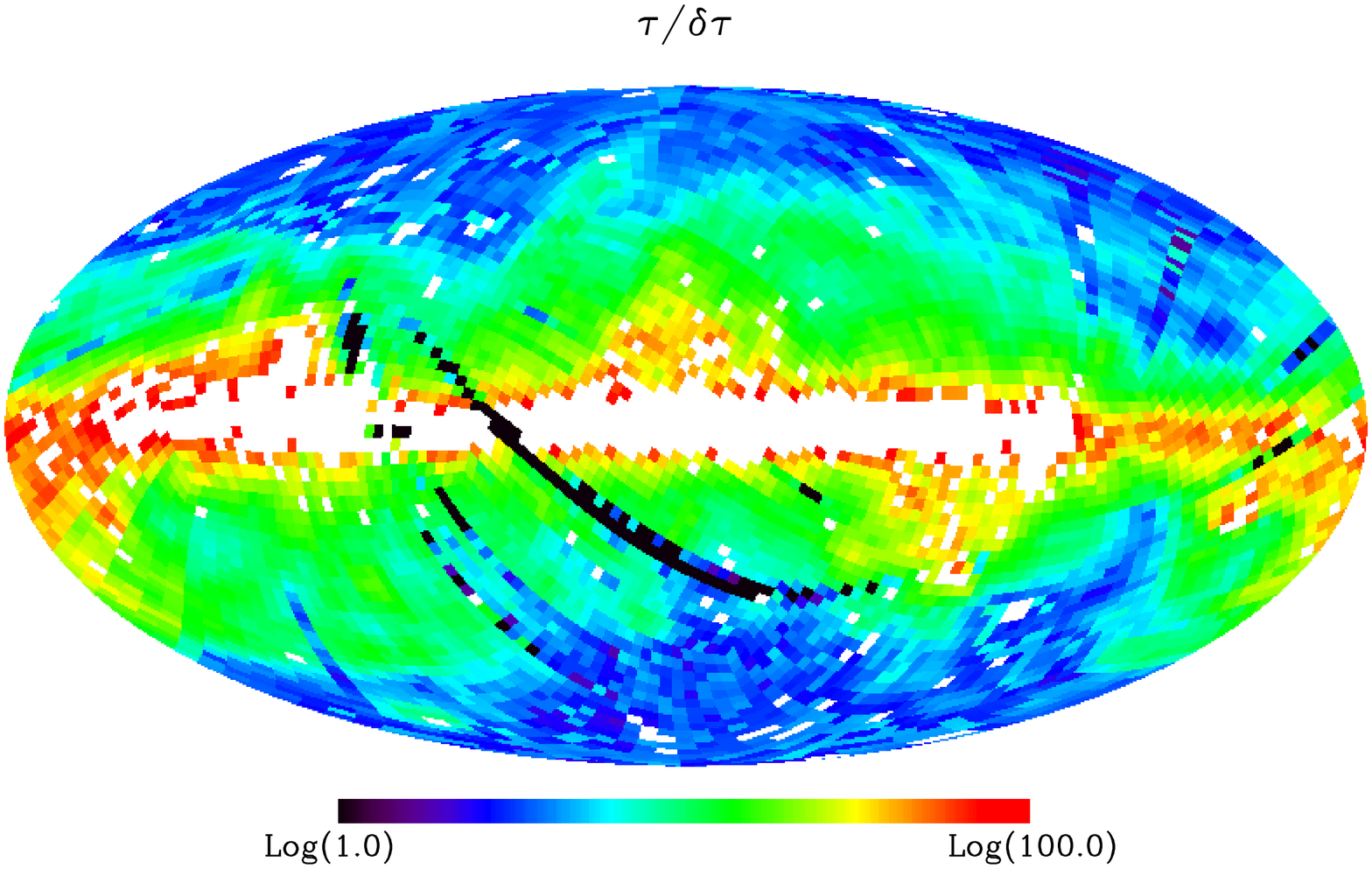}\\	
	\end{tabular}
	\caption{All-sky maps of $\chi^2_\mathrm{dof}$ (top left),  spatial resolution (top right), $T_\mathrm{dust}$ (second left), $T_\mathrm{dust}/\delta T_\mathrm{dust}$ (second right), $\tau$ (bottom left), and $\tau/\delta \tau$ (bottom right) of the one-component $\beta = 1.7$ fits that satisfy  $T_\mathrm{dust}/\delta T_\mathrm{dust} \ge 10.0$ (from \citealt{Liang11}). These maps are in Galactic coordinates Mollweide projection with the Galactic centre at the centre and longitude increasing to the left. The regional size map shows that majority of the fits are at the 6.71-$\sq^2$ level. In the parameter maps a pixel is masked in white if it corresponds to a $\chi^2$ with less than 10~per~cent probability. The group of black pixels that slant from the NEP to the SEP are positions where FIRAS did not provide data. }
	\label{fig:param_Tsn_ge20_alpha1p7}
\end{figure*}

A plot of $\chi^2_\mathrm{dof}$ distributions for 13 all-sky fixed-$\beta$ models with $\beta$ in the range $1.4-2.6$ at 0.1 increment and  $T_\mathrm{dust}/\delta T_\mathrm{dust} \ge 10$ is presented in Fig. \ref{fig:chi2_hist_all_alphas}. The high-$\chi^2$ tails of these distributions show that models with the largest (2.6) and smallest (1.4) values of $\beta$ have more fits with large $\chi^2_\mathrm{dof}$. The plot of $\chi^2_\mathrm{dof}$ excess for these all-sky models is presented in Fig. \ref{fig:chi2_hist_all_alphas}. It shows that the all-sky $\beta = 1.7$ model can fit the largest amount of data (87.6~per~cent area of the full sky). 
Fig. \ref{fig:param_Tsn_ge20_alpha1p7} presents sky maps of $\chi^2_\mathrm{dof}$, spatial resolution, dust temperature, optical depth, and the signal-to-noise of parameters from fitting the one-component $\beta = 1.7$ model. Each of the 6063 fits presented there satisfies the $T_\mathrm{dust}/\delta T_\mathrm{dust} \ge 10$ requirement with the least amount of spatial averaging. In the parameter maps, if a fit has less than 10~per~cent $\chi^2$ probability (i.e., $\chi^2_\mathrm{dof} > 1.13$), then its corresponding pixel is masked in white. Fit results for other $\beta$ models at different levels of constraint on $T_\mathrm{dust}/\delta T_\mathrm{dust}$ are provided in \citet{Liang11}.

\subsubsection{$\beta/\delta \beta$ constraint on a free-$\beta$ model}\label{sec:alpha_SN_constraint}

\begin{figure*}
	\centering
  {\bf One-component free-$\beta$ fits with $\beta/\delta \beta \ge 10.0$}	

	\begin{tabular}{cc}
	\hspace{1pc}\\
    \includegraphics[scale=0.31]{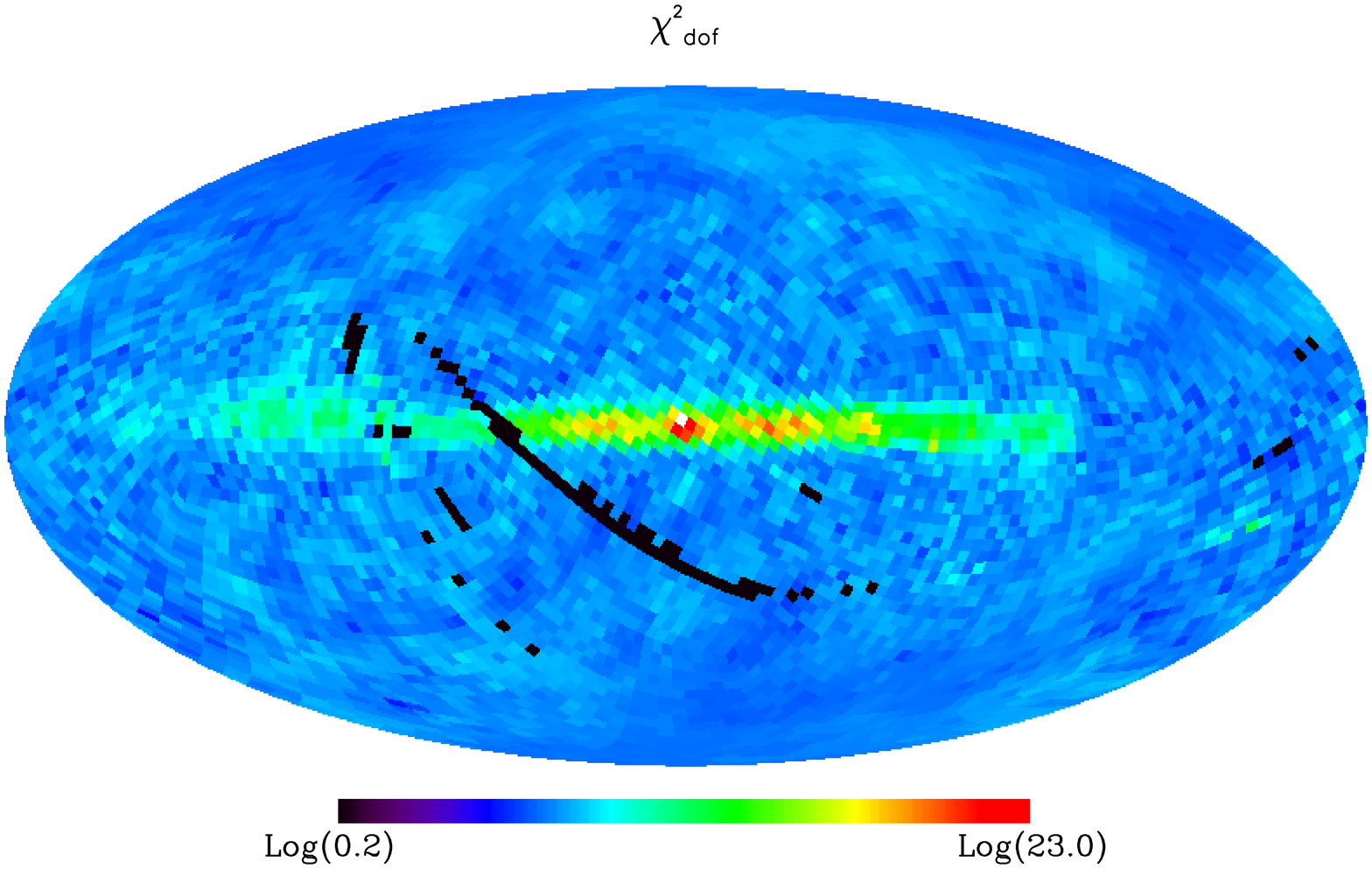}& 
    \includegraphics[scale=0.31]{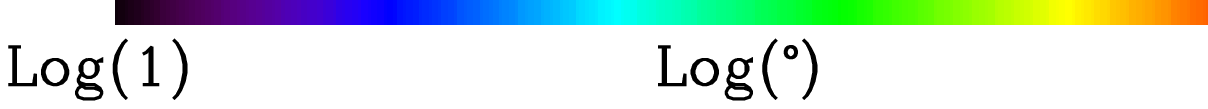}\\
	\hspace{1pc}\\
    \includegraphics[scale=0.31]{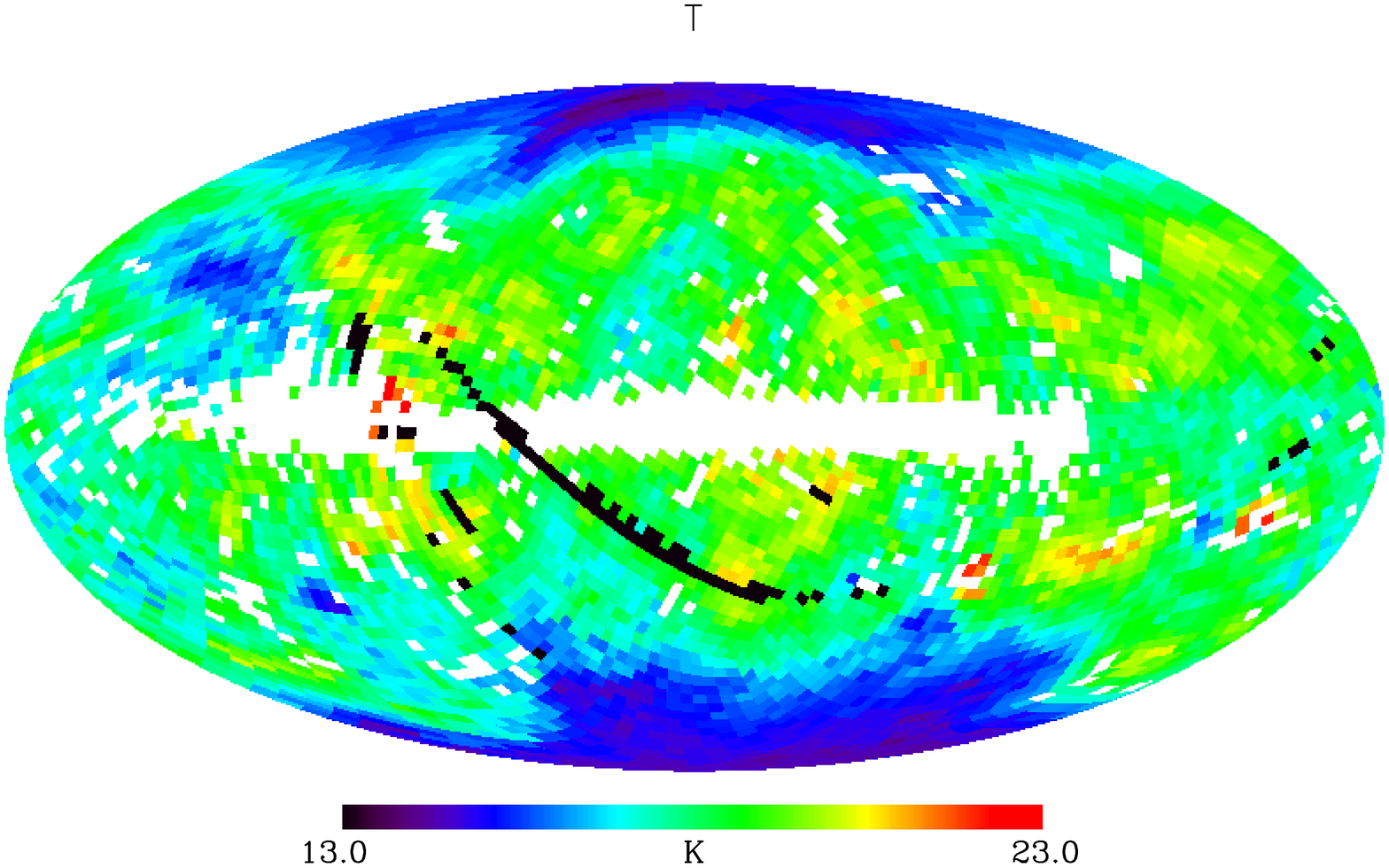}&
    \includegraphics[scale=0.31]{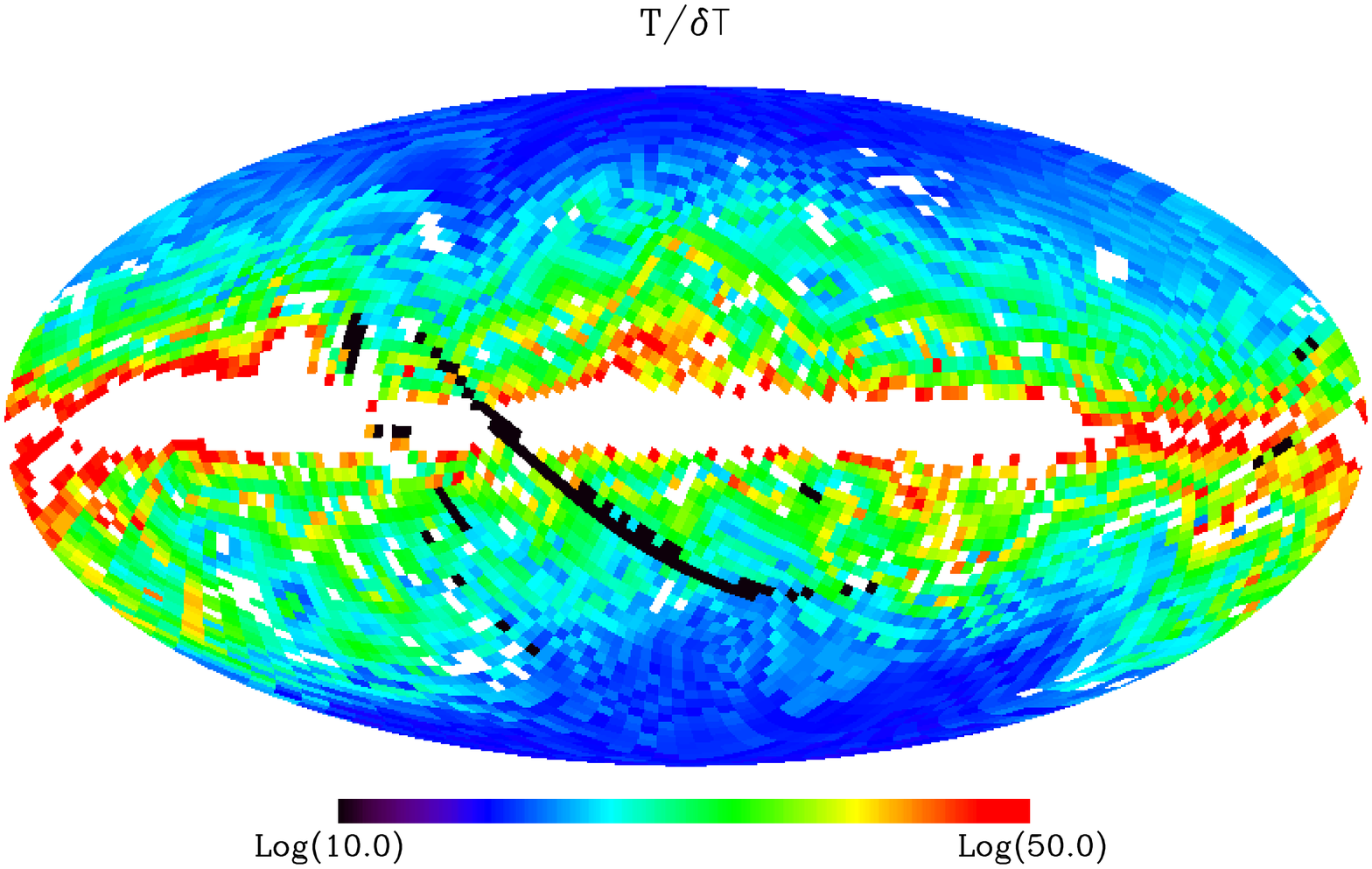}\\
	\hspace{1pc}\\
    \includegraphics[scale=0.31]{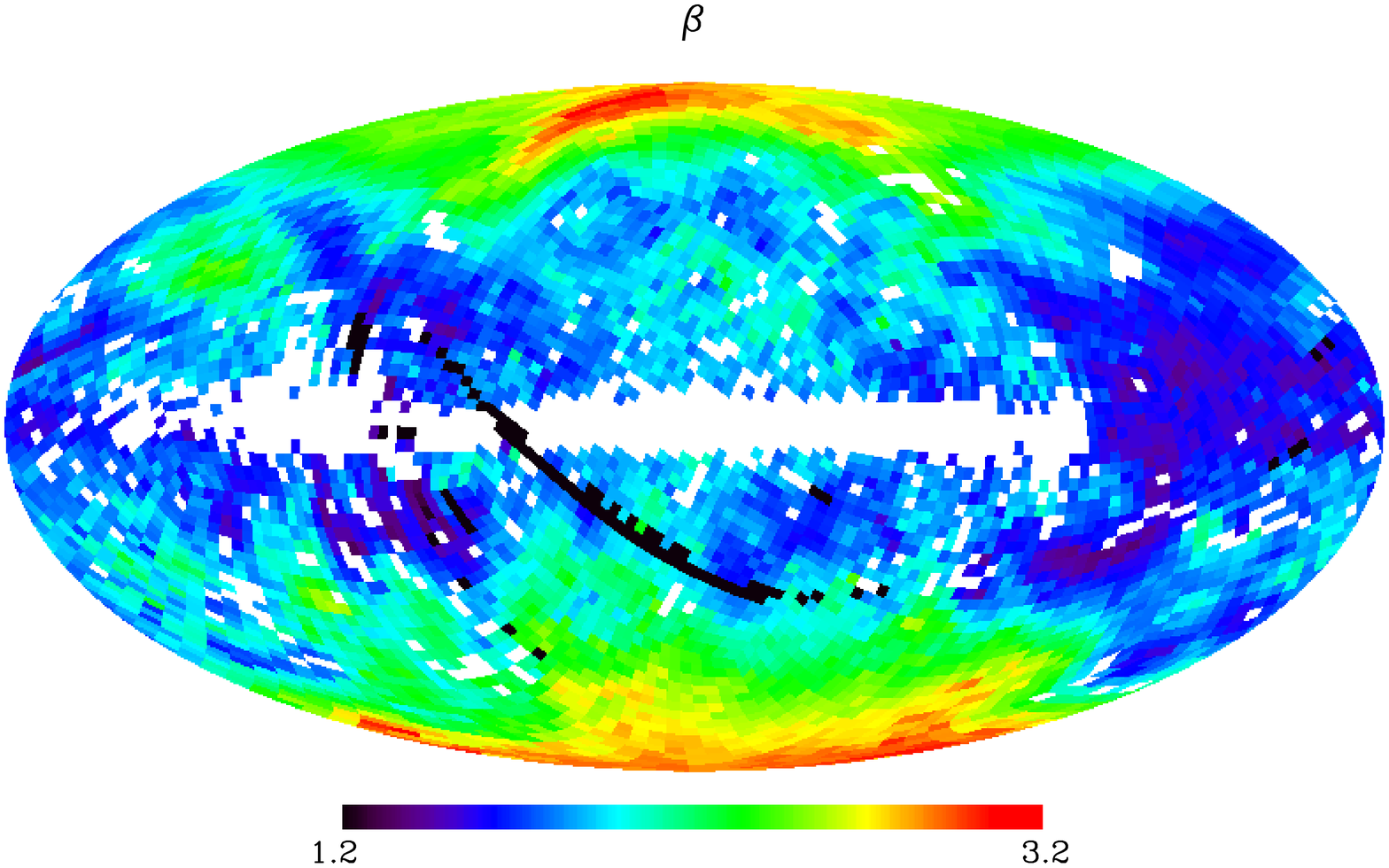}&	
    \includegraphics[scale=0.31]{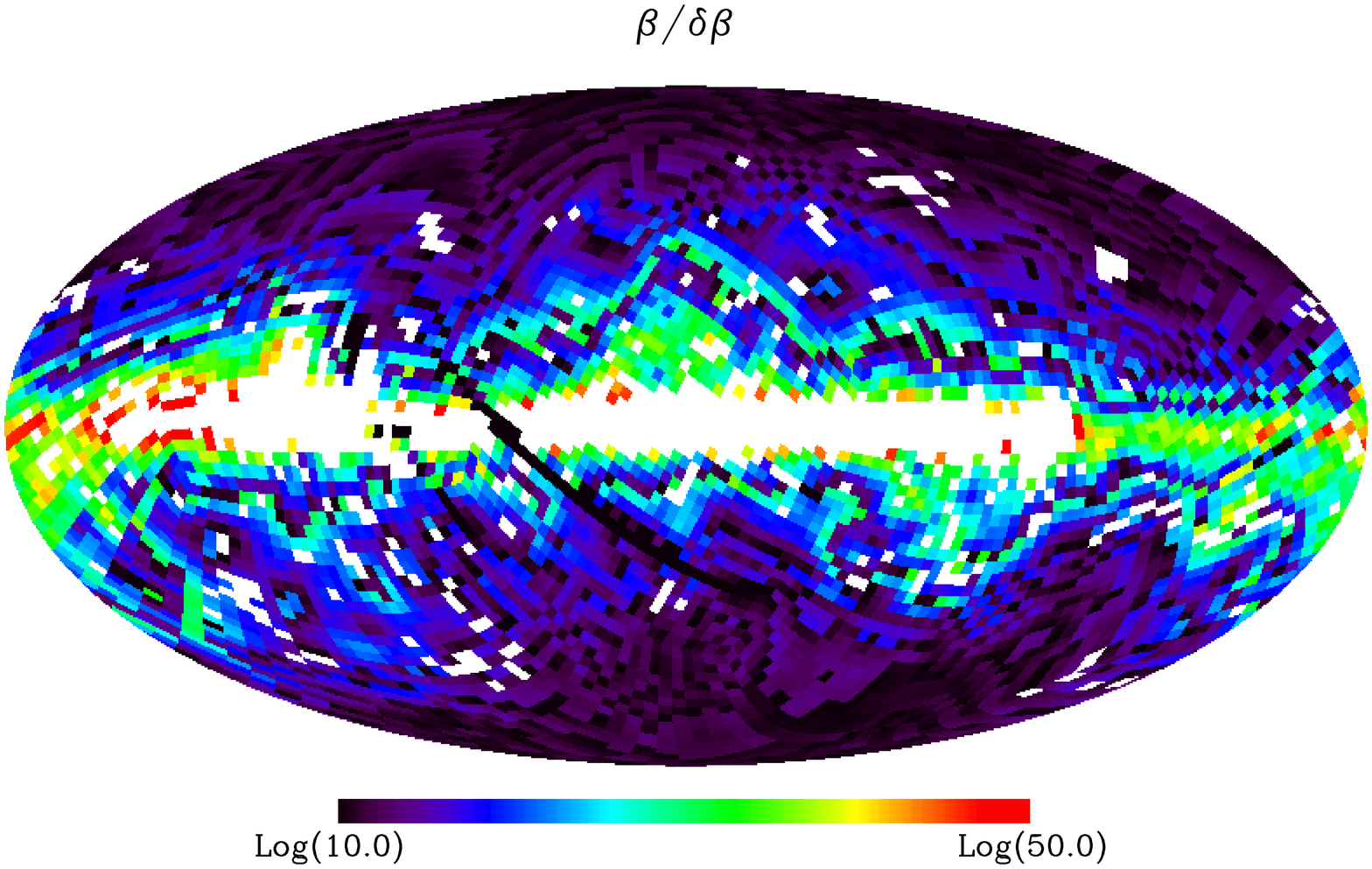}\\	
	\hspace{1pc}\\
    \includegraphics[scale=0.31]{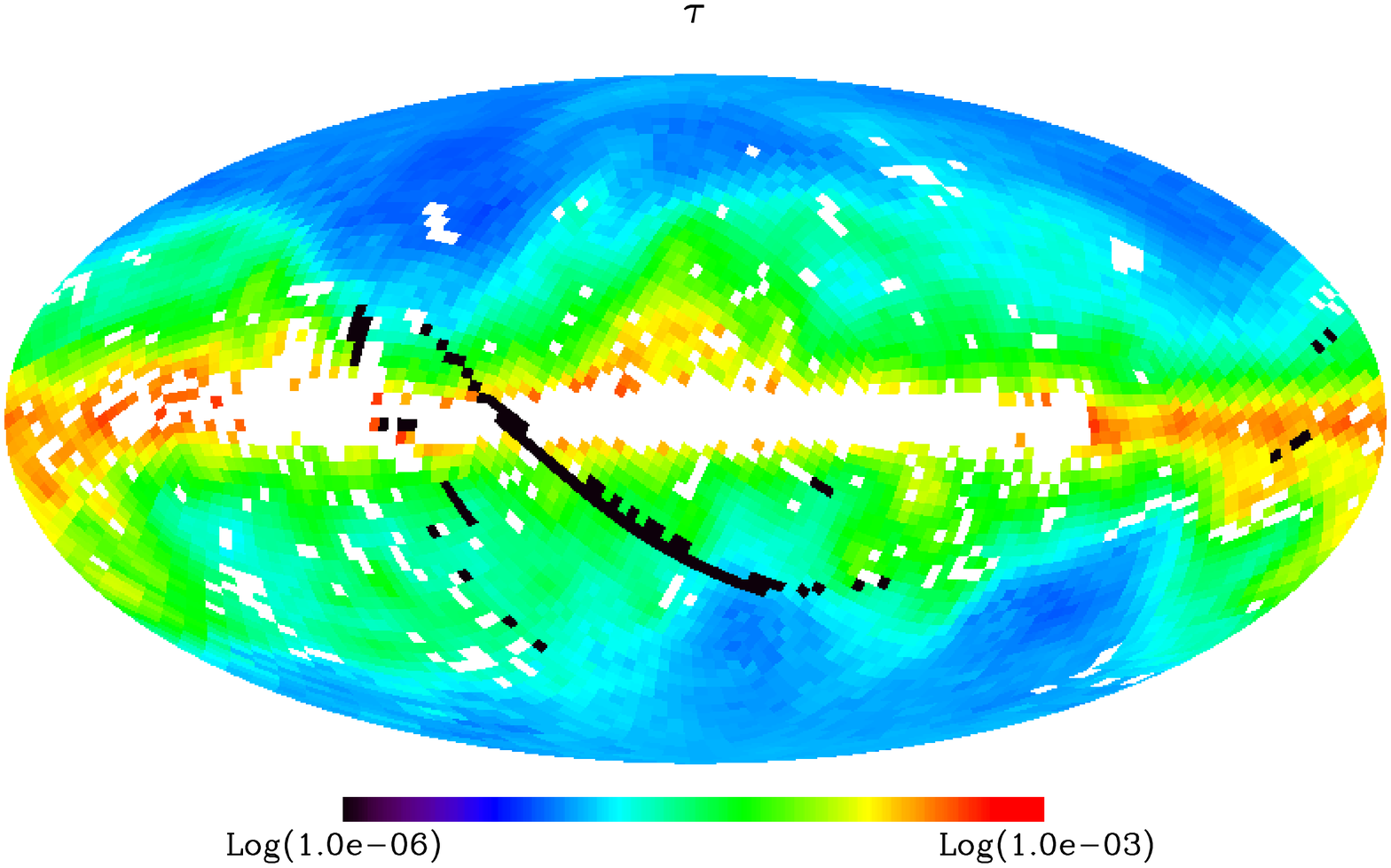}&	
    \includegraphics[scale=0.31]{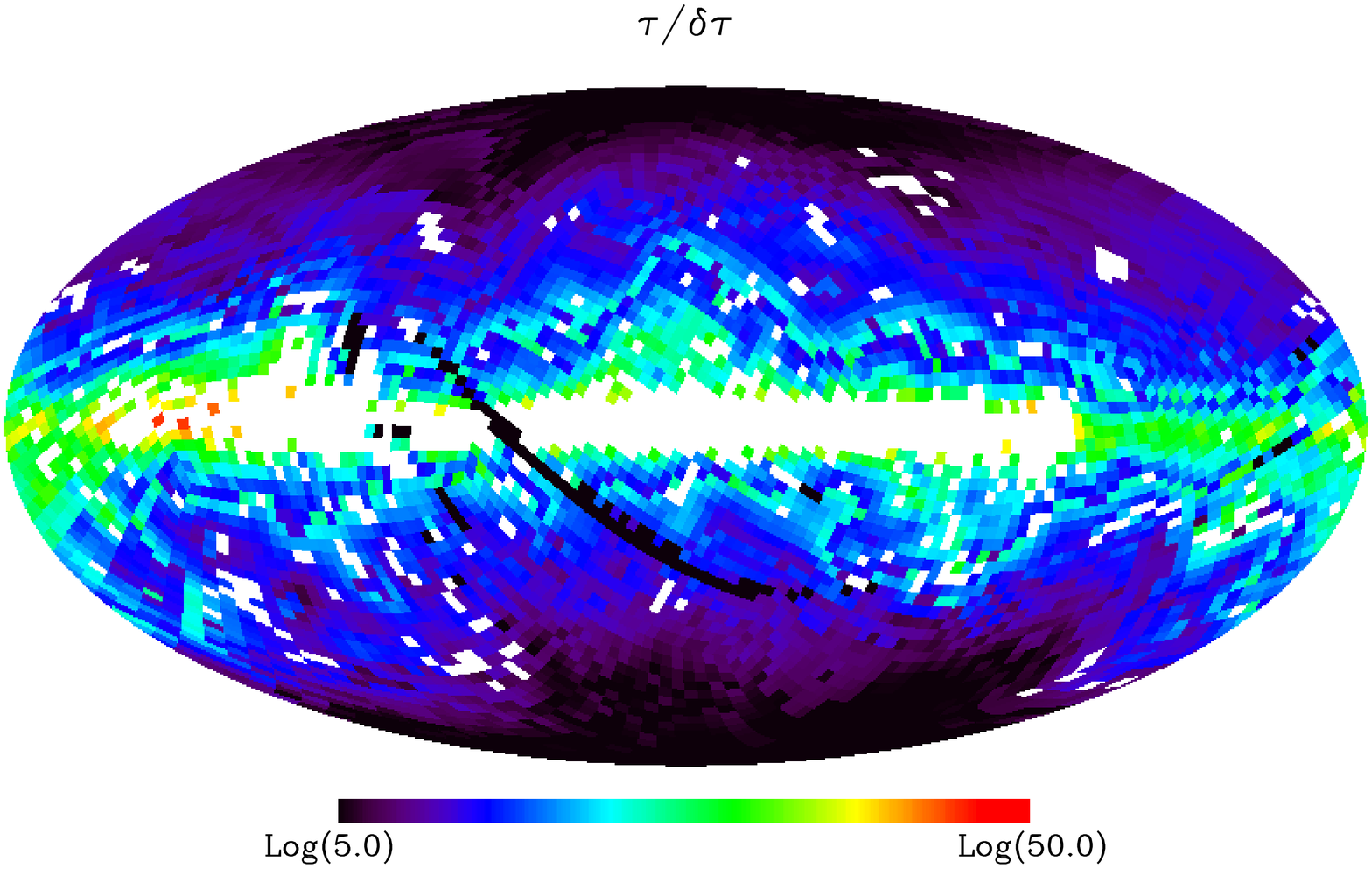}\\	
	\end{tabular}
	\caption{All-sky maps of $\chi^2_\mathrm{dof}$ (top left), spatial resolution (top right), dust temperature (second left), emissivity spectral index (third left), optical depth (bottom left), and their error estimates (right column) of the best-fitting one-component free-$\beta$ fits that satisfies $\beta/\delta \beta \ge 10.0$ in Galactic coordinates Mollweide projection with the Galactic centre at the centre and longitude increasing to the left. In the parameter maps a pixel is masked in white if it corresponds to a $\chi^2_\mathrm{dof} \ge 1.13$. The group of black pixels that slant from the NEP to the SEP are positions where FIRAS did not provide data. Notice that dust temperature predictions at the Galactic polar caps are noticeably lower than the rest of the sky, consistent with our knowledge that the Galactic poles have fewer and cooler stars. Warm regions on the temperature map correspond to known emission sources (see text).}
	\label{fig:param_free_alpha_var_res}
\end{figure*}

We apply a similar strategy to constrain fits that use a free-$\beta$ model. Instead of focusing on the signal-to-noise of the dust temperature, we now use signal-to-noise of the emissivity spectral index to gauge the amount of spectral averaging. As an example, sky maps of the best-fitting parameters and their signal-to-noise for the free-$\beta$ model with $\beta / \delta \beta \ge 10.0$ are presented in Fig. \ref{fig:param_free_alpha_var_res}.

\begin{table*}
\begin{minipage}{140mm}
\centering
\caption{Spatial resolution of the one-component free-$\beta$ fits with $\chi^2_\mathrm{dof} \le 1.13$}
\begin{tabular}{|c|c|c|c|c|} 
\hline\\
Level & Regional size of fits & \multicolumn{3}{|c|}{Percentage of the full sky at a regional average} \\
 & \multicolumn{1}{|c|}{($\sq^2$)} & $\beta / \delta \beta \ge 5.0$ & $\beta / \delta \beta \ge 6.7$ & $\beta / \delta \beta \ge 10.0$ \\		
\hline\\
~1 & ~~~6.71 & 40.01 & 30.63 & 22.09 \\		%
~2 & ~~60.43 & 23.14 & 24.53 & 22.51 \\
~3 & ~167.86 & 11.87 & 11.07 & 10.60 \\
~4 & ~329.00 & ~4.61 & ~8.59 & ~6.41 \\
~5 & ~543.86 & ~3.53 & ~3.48 & ~5.53 \\
~6 & ~812.44 & ~2.18 & ~2.78 & ~4.62 \\
~7 & 1134.73 & ~1.01 & ~2.12 & ~3.30 \\
~8 & 1510.73 & & ~1.92 & ~2.59 \\
~9 & 1940.45 & & ~1.22 & ~2.23 \\
10 & 2423.88 & & & ~2.08 \\
11 & 2961.03 & & & ~1.90 \\
12 & 3551.89 & & & ~1.53 \\
13 & 4196.47 & & & ~0.72 \\
14 & 4894.76 & & & ~0.23 \\
\hline\\
 & Total & 87.66 & 87.19 & 86.34 \\
\hline
\end{tabular}
\label{tbl:npix_free_alpha}\\
\end{minipage}
\end{table*}

Notice that with spatial averaging the dust temperature map has more consistent values at high latitudes near the Galactic poles than that obtained from the 7$^\circ$-pixel fits in Fig. \ref{fig:discrete_alpha_pixel_fit_skymaps}. The uncertainty of $T_\mathrm{dust}$ is less than 7.4~per~cent as a result of the constraint on $\beta / \delta \beta$, and the uncertainty of $\tau$ has a maximum of 23.25~per~cent. The spatial resolution of these fits are presented in an all-sky map in Fig. \ref{fig:param_free_alpha_var_res} and summarized in Table \ref{tbl:npix_free_alpha}.

\begin{figure}
	\centering
      	\includegraphics[scale=0.55]{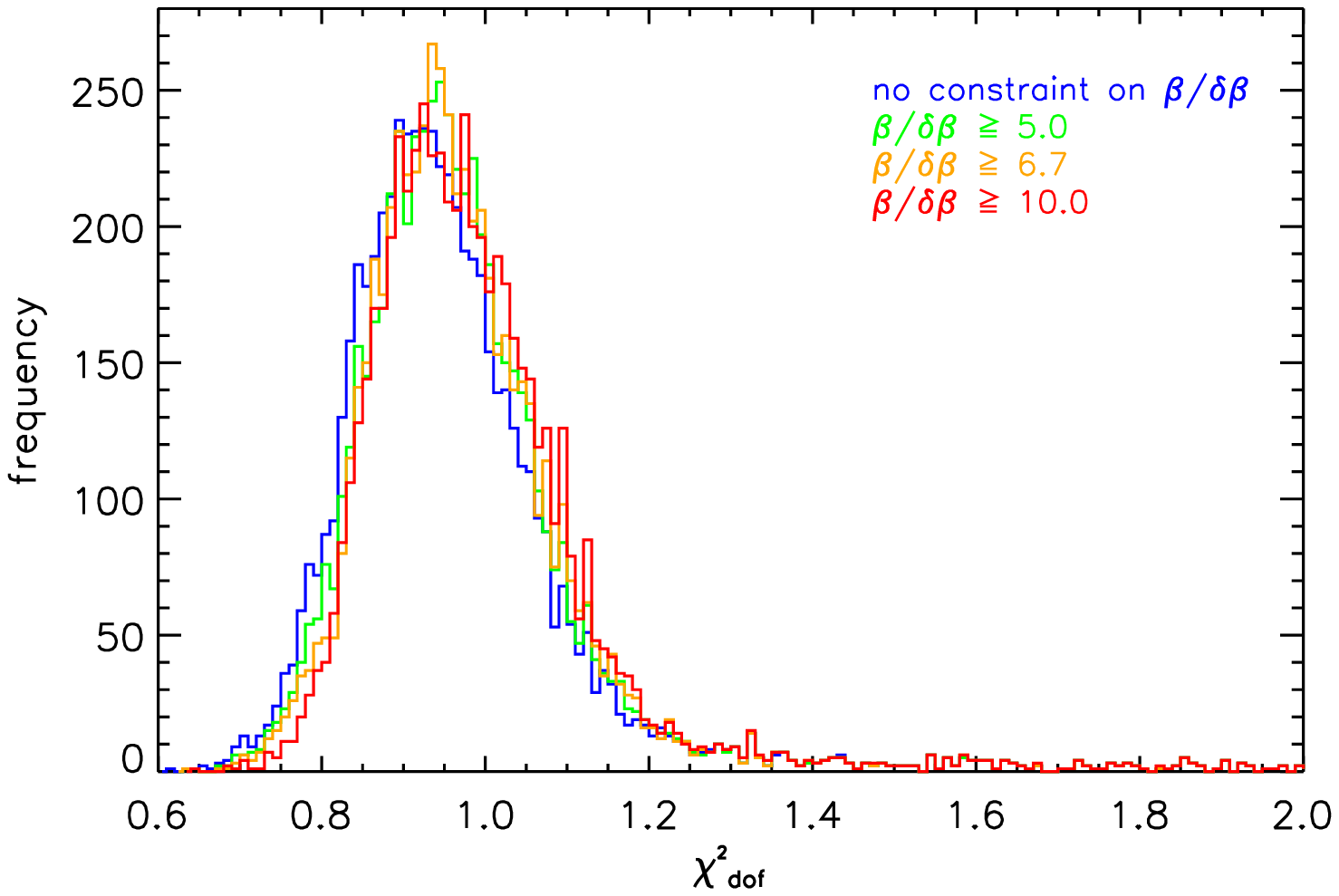}\\
		\hspace{2pc}\\
      	\includegraphics[scale=0.55]{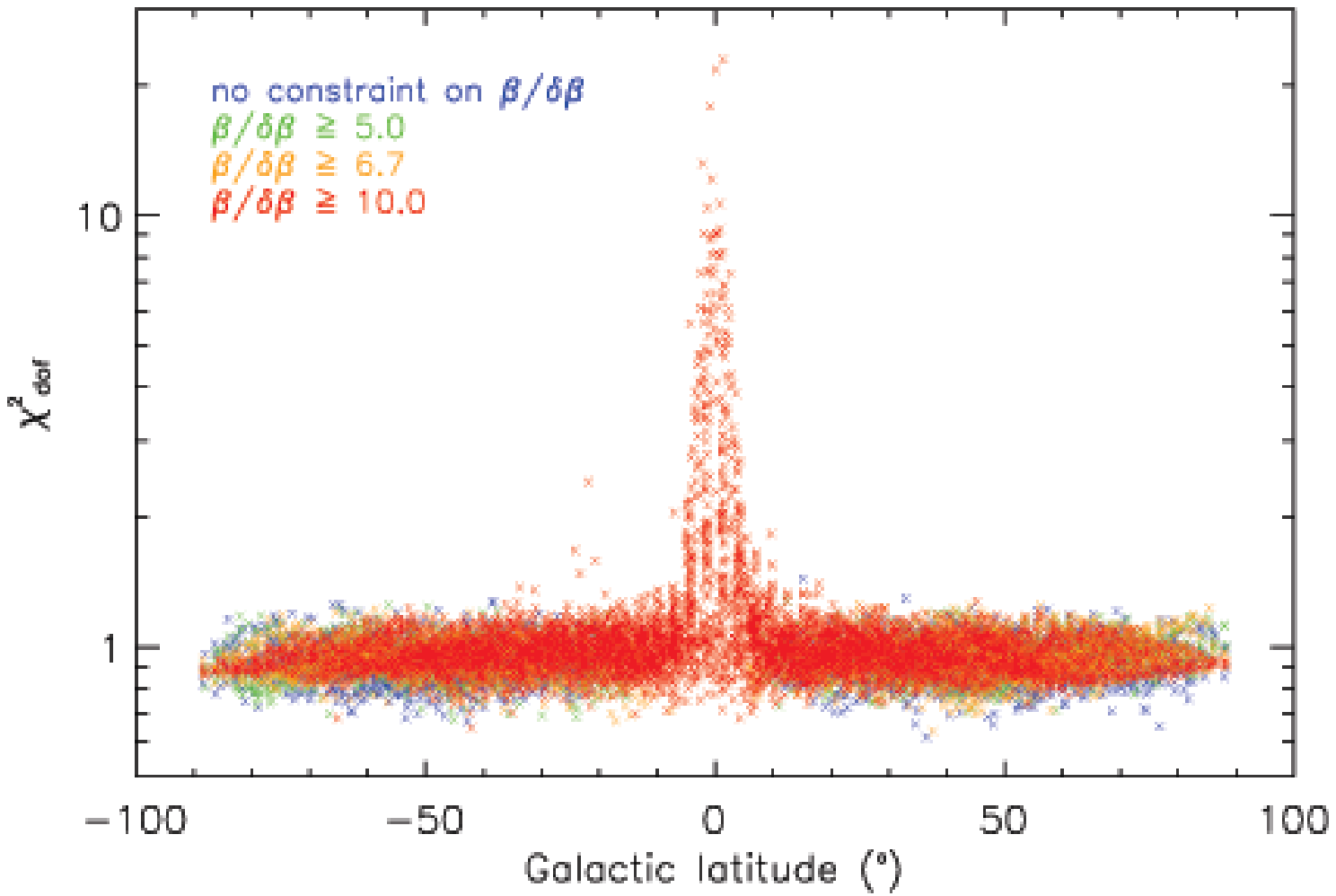}\\
	\caption{Upper: Distributions of $\chi^2_\mathrm{dof}$ from one-component free-$\beta$ fits that use no constraint on any parameter and those that satisfy $\beta /\delta \beta \ge$ 5.0, 6.7 and 10.0, respectively. Lower: $\chi^2_\mathrm{dof}$ vs. Galactic latitude.}
	\label{fig:chi_hist_and_vs_glat_free_alpha}
\end{figure}

The upper panel in Fig. \ref{fig:chi_hist_and_vs_glat_free_alpha} compares $\chi^2_\mathrm{dof}$ distributions of fits that use no constraint on any parameter and those that satisfy $\beta /\delta \beta \ge$ 5.0, 6.7, and 10.0. The shape of the $\chi^2_\mathrm{dof}$ distributions again resemble a Gaussian, and the constrained distributions all centre around 0.95. This means that there is no apparent systematic bias in the fits and the error estimates for the data is about right.

The plot of $\chi^2_\mathrm{dof}$ vs. Galactic latitude, lower panel of Fig. \ref{fig:chi_hist_and_vs_glat_free_alpha}, shows that at $|b| < 10^\circ$, $\chi^2_\mathrm{dof}$ continues to be $> 1$ as is the case of fitting all-sky one-component fixed-$\beta$ models. At high latitudes, $\chi^2_\mathrm{dof}$ do not flare up with increasing constraint on $\beta / \delta \beta$, as oppose to that which happens when $T_\mathrm{dust} / \delta T_\mathrm{dust}$ requirements are imposed on a fixed-$\beta$ model.

\begin{figure}
	\centering
      	\includegraphics[scale=0.55]{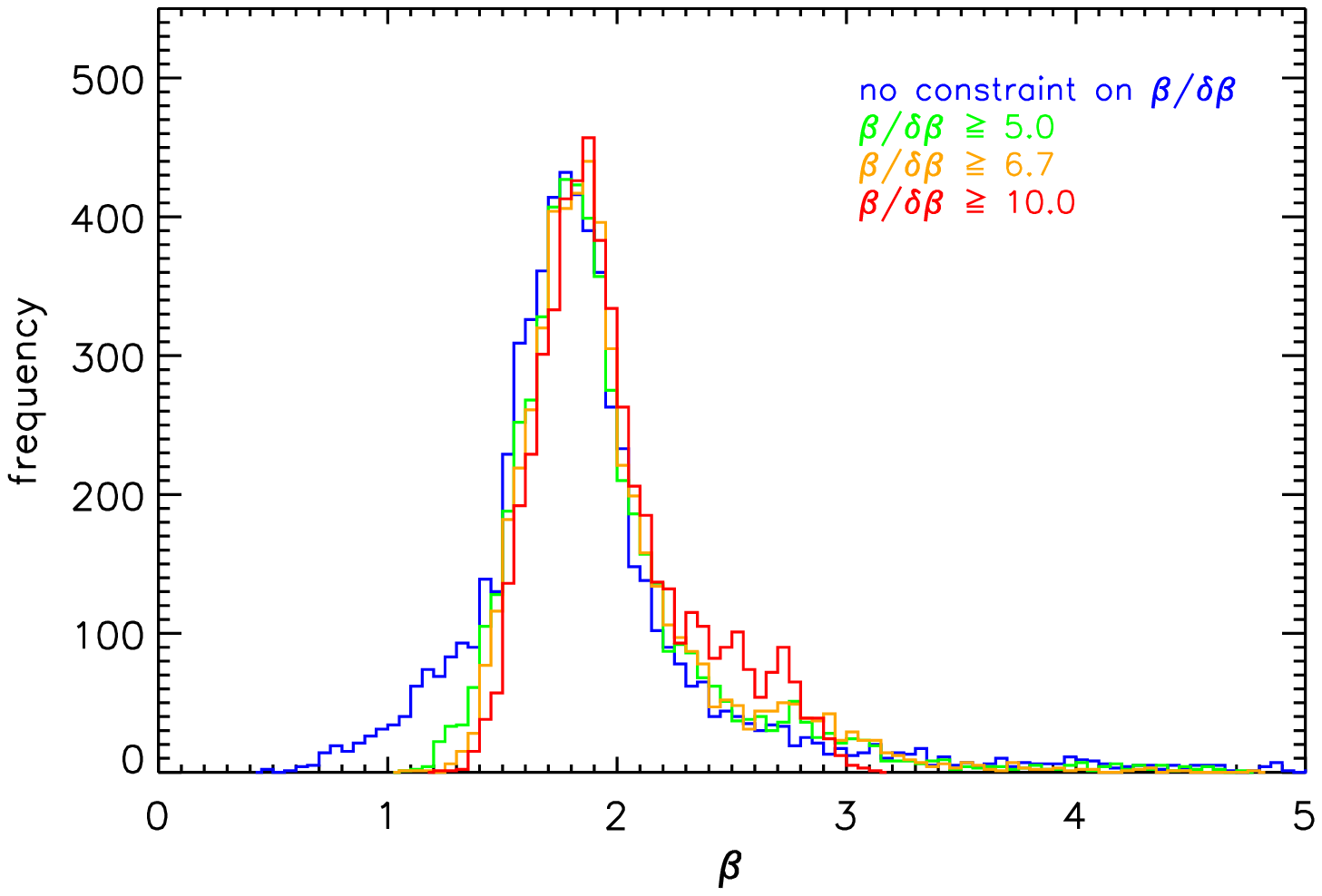}\\
		\hspace{2pc}\\
      	\includegraphics[scale=0.55]{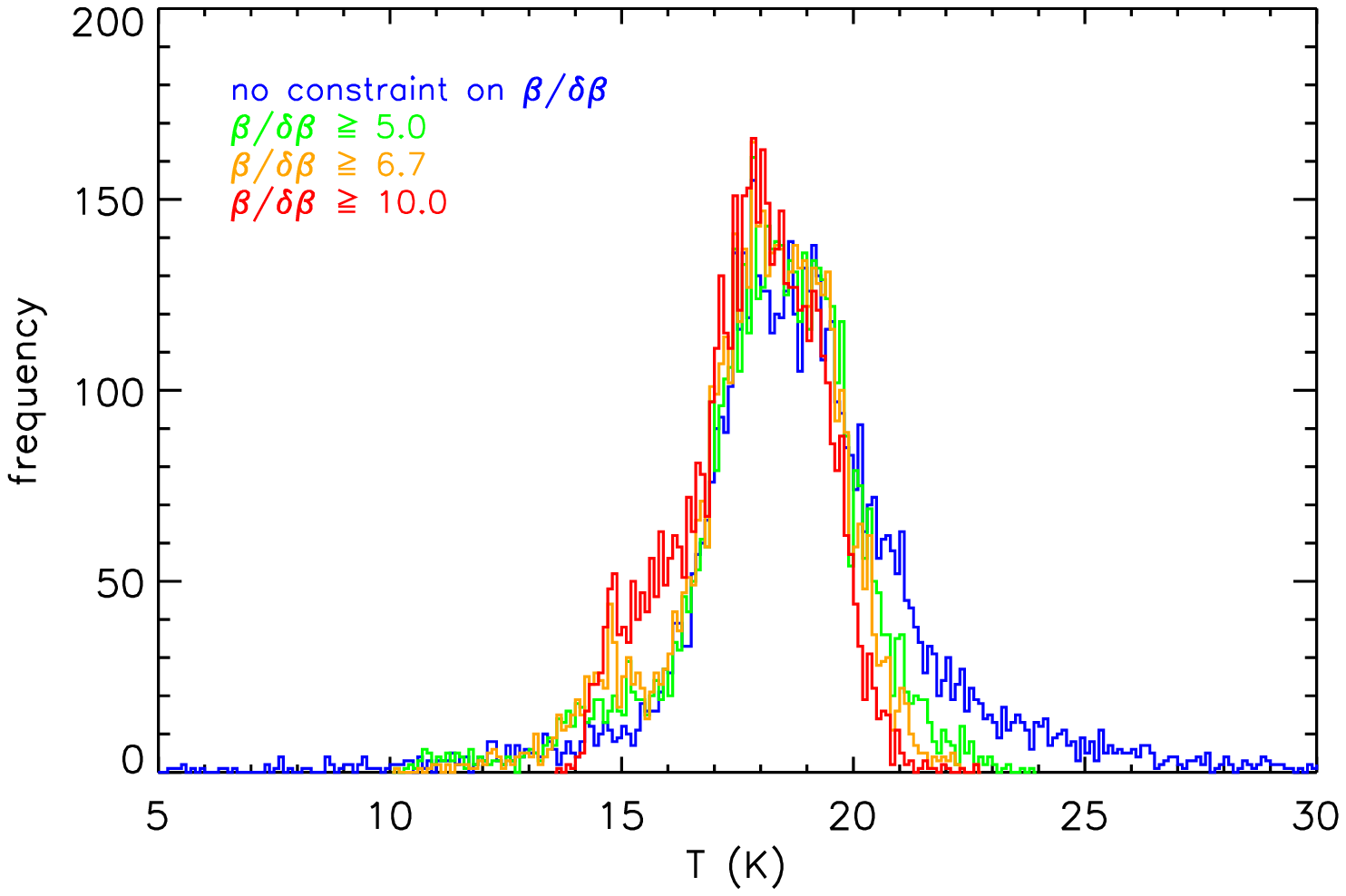}\\
	\caption{Distributions of $\beta$ (top) and $T_\mathrm{dust}$ (bottom) 
of one-component free-$\beta$ fits that use no constraint on any parameter and those that satisfy $\beta /\delta \beta \ge$ 5.0, 6.7 and 10.0, respectively. Only those fits with $\chi^2_\mathrm{dof} \le 1.13$ are included.}
	\label{fig:alpha_T_hist_free_alpha}
\end{figure}

Fig. \ref{fig:alpha_T_hist_free_alpha} presents distributions of $\beta$ and $T_\mathrm{dust}$ of one-component free-$\beta$ fits that use no constraint on any parameter and those that satisfy $\beta /\delta \beta \ge$ 5.0, 6.7, and 10.0. The centres of the distributions are at 1.80, 1.83, 1.85, and 1.88, respectively. This plot shows that the $\beta / \delta \beta$ requirement has the effect of moving $\beta$ from below 1.5 to higher values. 
With even a moderate amount of constraint on $\beta / \delta \beta$, the range of $\beta$ values quickly reduces to between 1 and 3, an indication that values outside this range are rare in nature. 

Also shown in Fig. \ref{fig:alpha_T_hist_free_alpha}, the $T_\mathrm{dust}$ distributions do not peak at a single value. Instead, there is a range of most popular temperatures between 17 and 20 K. Compared to the unconstrained case, the $\beta / \delta \beta$ requirement smooths out the high-temperature points and effectively replaces them with values at or below 20 K.

\begin{table*}
\begin{minipage}{160mm}
\centering
\caption{Parameters and their uncertainties of the one-component free-$\beta$ fits with $\chi^2_\mathrm{dof} \le 1.13$}
\begin{tabular}{|l|c|c|c|} 
\hline\\
Constraint on $\beta / \delta \beta$ & \multicolumn{2}{|c|}{$T_\mathrm{dust}$} & $\delta T_\mathrm{dust}$ \\
 & \multicolumn{2}{|c|}{(K)} & (K) \\
 & min & max & \\		
\hline\\
$\ge ~5.0$ & 10.11 & 23.83 & 2.67 \\		%
$\ge ~6.7$ & 10.12 & 22.69 & 1.95 \\
$\ge 10.0$ & 13.69 & 22.69 & 1.26 \\
\hline\\
\hspace{3pc}\\
\hline\\
Constraint on $\beta / \delta \beta$ & \multicolumn{2}{|c|}{$\beta$} & $\delta \beta$ \\
 & min & max &  \\		
\hline\\
$\ge ~5.0$ & 1.08 & 4.71 & 0.91\\		%
$\ge ~6.7$ & 1.08 & 4.80 & 0.71\\
$\ge 10.0$ & 1.24 & 3.13 & 0.31\\
\hline\\
\hspace{3pc}\\
\hline\\
Constraint on $\beta / \delta \beta$ & \multicolumn{2}{|c|}{$\tau$} & $\delta \tau / \tau$ \\
 & \multicolumn{2}{|c|}{$\times 10^{-5}$} & (per cent) \\
 & min & max & \\		
\hline\\
$\ge ~5.0$ & 0.33 & 46.15 & 62.83 \\		%
$\ge ~6.7$ & 0.45 & 46.15 & 49.02 \\
$\ge 10.0$ & 0.61 & 46.15 & 23.25 \\
\hline\\
\end{tabular}
\label{tbl:param_free_alpha}\\
\end{minipage}
\end{table*}

The amounts of constraint on dust temperature, optical depth and emissivity spectral index for various levels of constraint on $\beta / \delta \beta$ are summarized in Table \ref{tbl:param_free_alpha}. In order to place adequate constraint on each parameter and to keep the regional sizes low, we adopt the $\beta / \delta \beta \ge 10.0$ constraint as our standard.

%% file: fixed_vs_free_alpha_dust6.tex
\subsection{Dust temperature at high Galactic latitudes}

Dust temperature is not measured directly in astronomical observations. Instead, one derives dust temperature from intensity measurements of dust emission and a model that describes the emission process. That dust temperature decreases with the increase in Galactic latitude is supported by observations which identify thermal dust to be the origin of the Galaxy's diffuse emission, observations which show that the diffuse emission decreases from low to high Galactic latitudes following a cosecant law, observations of the cosecant law for dust reddening and extinction in the Galaxy with respect to Galactic latitude, and observations of the decreasing number of bright stars toward the Galactic polar caps. Substantial amount of work has been carried out over the last four decades, we mention only a few as examples: 

In reviewing global properties of dust in the Galaxy, \citet{Stein83} point out that ``the far-infrared luminosity in the Galaxy is associated with regions where stars are now or were recently forming,'' and so ``even in the galactic center region, much of the far-infrared luminosity is associated with ongoing or recent star formation activity.'' Using far-infrared luminosity as a measure of the current rate of star formation, dust acts as ``a frequency converter which absorbs the short wavelength photons emitted by the newly formed stars and re-emitting the energy at far-infrared wavelengths.''

That dust grains are heated by the general interstellar radiation field (ISRF) which follows a cosecant law to the Galactic poles and has a stellar origin have been confirmed by many research groups. For instance, by fitting spectra of diffuse emission from the Galactic plane \citet{Mezger82} and \citet{Mathis83} identify two main contributors to the diffuse emission at wavelengths $\ge 20$ $\mu$m: dust heated by O stars (60\%) and dust associated with diffuse atomic intercloud gas heated by the general interstellar radiation field (40\%). They conclude that dust associated with clouds containing no luminous sources of heating such as OB stars contributes less than 7\% of the total diffuse far-infrared/sub-millimeter emission. 
In 1986, Cox, Kr\"{u}egel \& Mezger showed that the total luminosity of the re-radiated dust emission from the Galactic disk is about 40\% of the total stellar luminosity. Their best-fitting dust model predicts that $\sim $ 90\% of the total Galactic infrared luminosity come from three components: a cold dust ($T \sim 15-25$ K) component associated with atomic hydrogen, a very cold dust ($<T> \sim 14$ K) component associated with molecular hydrogen and located inside quiescent molecular clouds, and a warm dust ($T \sim 30-40$ K) component associated with ionized gas in extended low density H{\sevensize II} regions heated by O and B stars. The first two components are heated by the general ISRF generated by both young and old stellar populations. They dominate the sub-millimeter part of the spectrum and have a total luminosity of $\sim 37\%$ of the total Galactic infrared luminosity. The luminosity of the warm dust component, on the other hand, accounts for $\sim 50\%$ of the total Galactic IR luminosity. 
Using the first all-sky infrared measurements taken by the {\sl Infrared Astronomical Satellite} ({\sl IRAS}, \citet{IRAS88}), \citet{Boulanger88} show that Galactic emission exists at Galactic latitude $|b|>10^\circ$ at 12, 25, 60 and 100 $\mu$m and that the emission comes mainly from dust heated by the ISRF associated with atomic gas and diffuse ionized gas. More specifically, they show that the Galactic latitude profiles of the {\sl IRAS} all-sky measurements at 60 and 100 $\mu$m follow a cosecant law from 10$^\circ$ to the poles. This means that dust emission decreases from low to high latitudes in the manner as would be expected for a plane-parallel configuration.

Not only does the strength of the ISRF decreases with increasing latitude, but the amount of dust at high Galactic latitudes also decreases following a cosecant law. 
Distribution of dust at high Galactic latitudes can be inferred from studies of dust reddening and extinction. For example, \citet{McClure71} use 207 late-type stars (K giants) near the north Galactic pole to calculate a reddening law and find that it is consistent with the cosecant law ($0.06 \cdot \mathrm{csc}~b$), where the Sun is assumed to be in a cylindrical hole of an absorbing layer in the Galactic plane. They conclude that there is almost no absorption from the pole down to 50$^\circ$. 
\citet{deVaucouleurs83} studies high-latitude extinction in extragalactic objects using five independent methods: counts of faint galaxies, counts of distant galaxy clusters, mean surface brightness of bright galaxies, hydrogen-luminosity ratio measured by the H{\sevensize I} index for bright galaxies, and total color indices of bright galaxies. They show that extinction derived from each of these methods supports the cosecant law from $|b|>20^\circ$ to the poles in each Galactic hemisphere.

Finally, we know that stars are the power houses in the Galaxy for both the ISRF and interstellar dust grains and that all-sky surveys have shown that the Galactic polar regions have fewer and cooler stars (sky maps in \citealt{IRAS88,Arendt98,Voges99,Gudennavar12}). With the observations of less intense emission, fewer dust particles and fewer sources of energy at the Galactic polar regions, we thus expect that dust temperature in those regions to be lower accordingly.

\subsection{Comparing dust temperature predictions of different models}

The crucial test to any model is comparison with observations. In this section, we present all-sky maps of dust temperature obtained from the literature and those from our study. We argue that since dust temperature distribution should generally decrease with the increase in Galactic latitude, models that fail such a test are unphysical.

\begin{figure*}
	\centering
	\begin{tabular}{cc}
      	\includegraphics[scale=0.31]{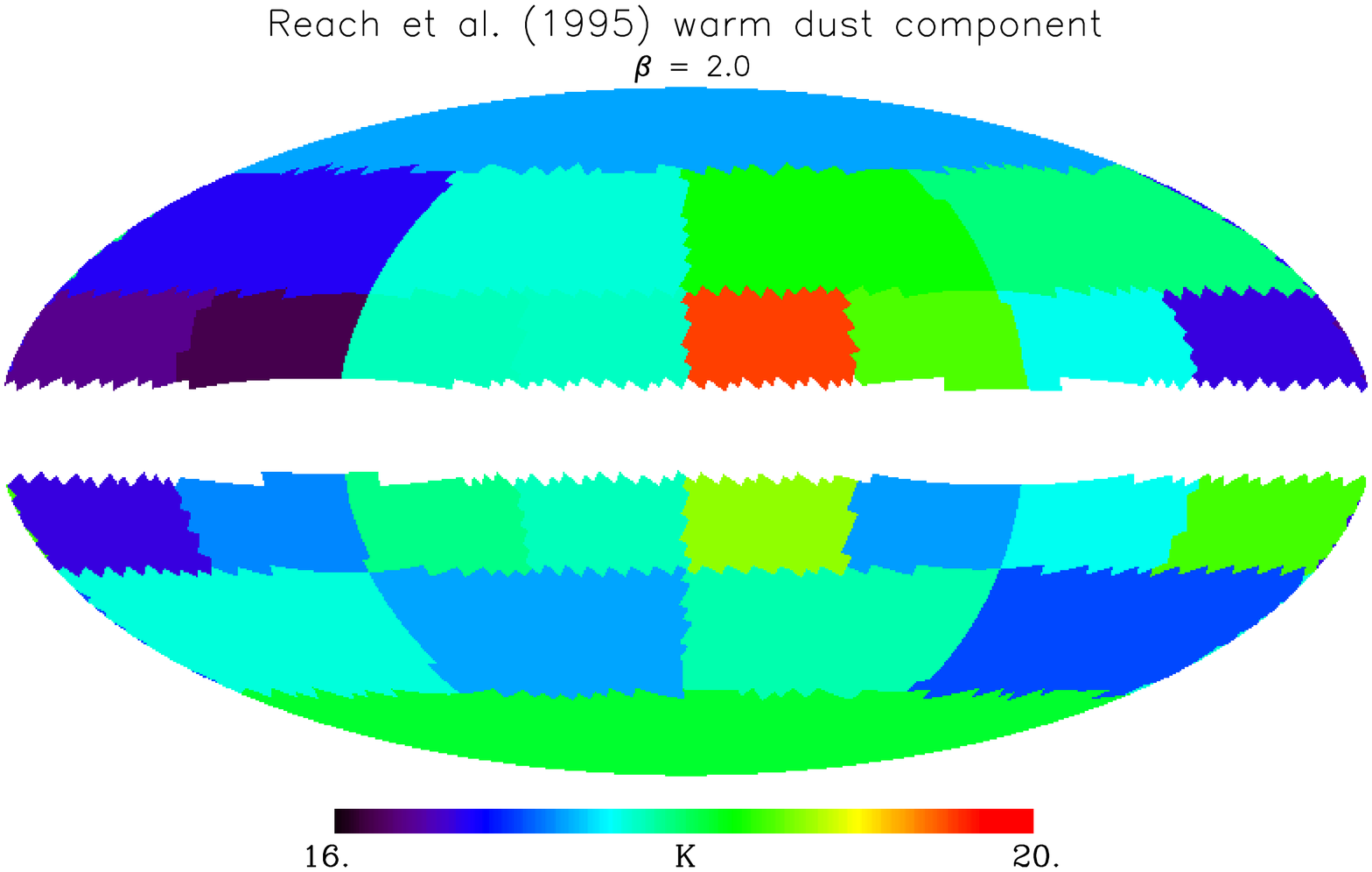}&	
      	\includegraphics[scale=0.31]{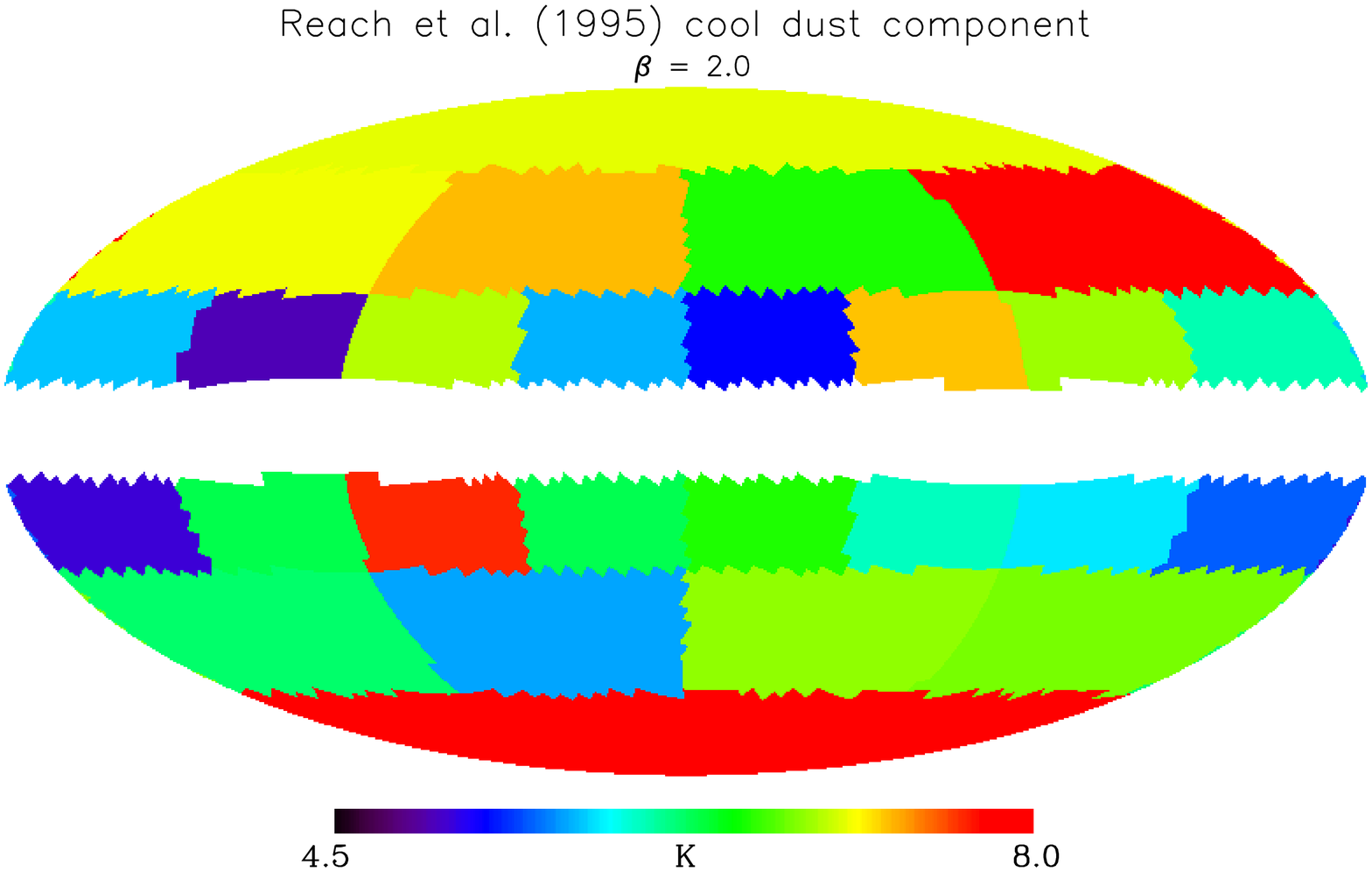}\\
				\hspace{1pc}\\
      	\includegraphics[scale=0.31]{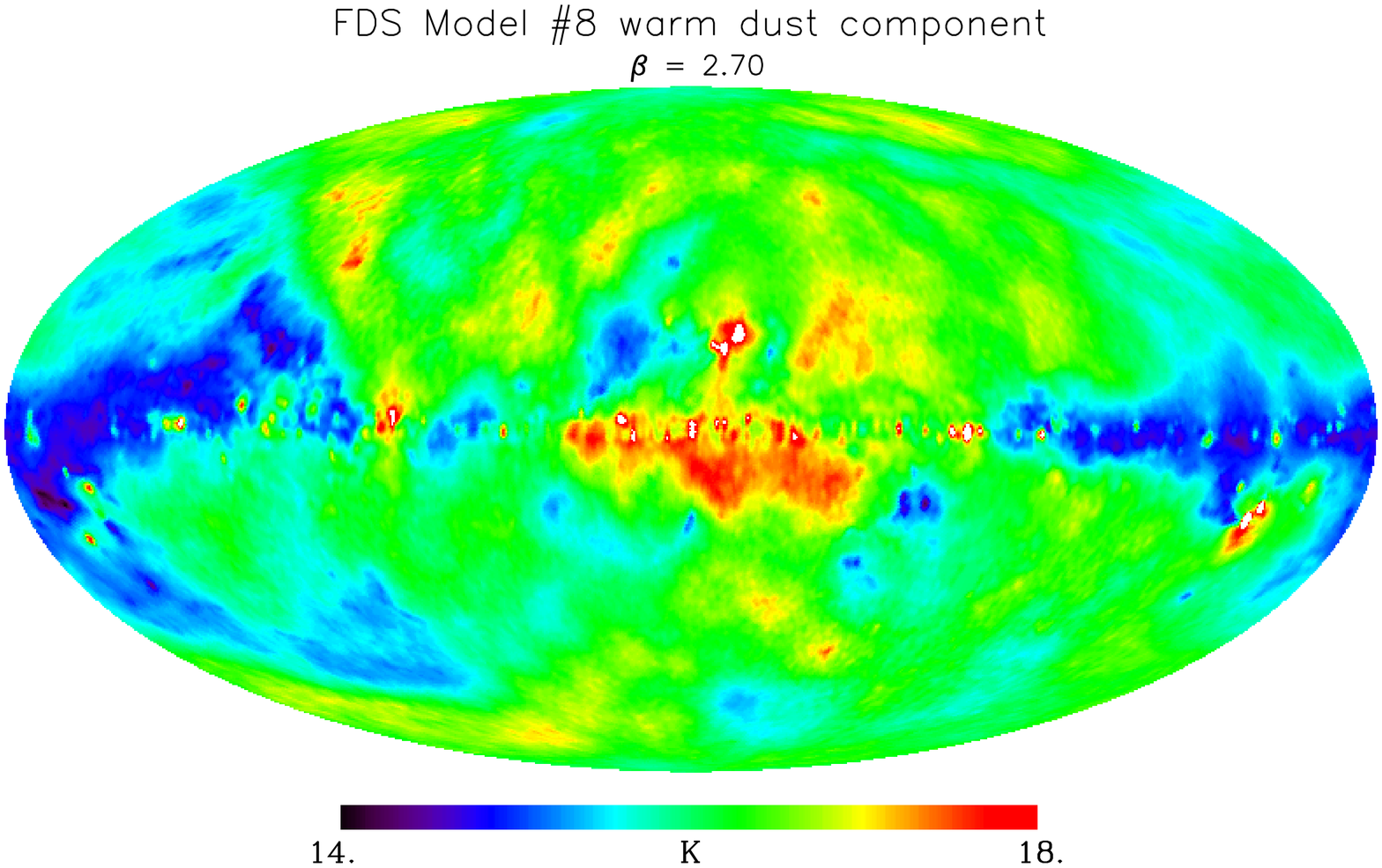}&	
      	\includegraphics[scale=0.31]{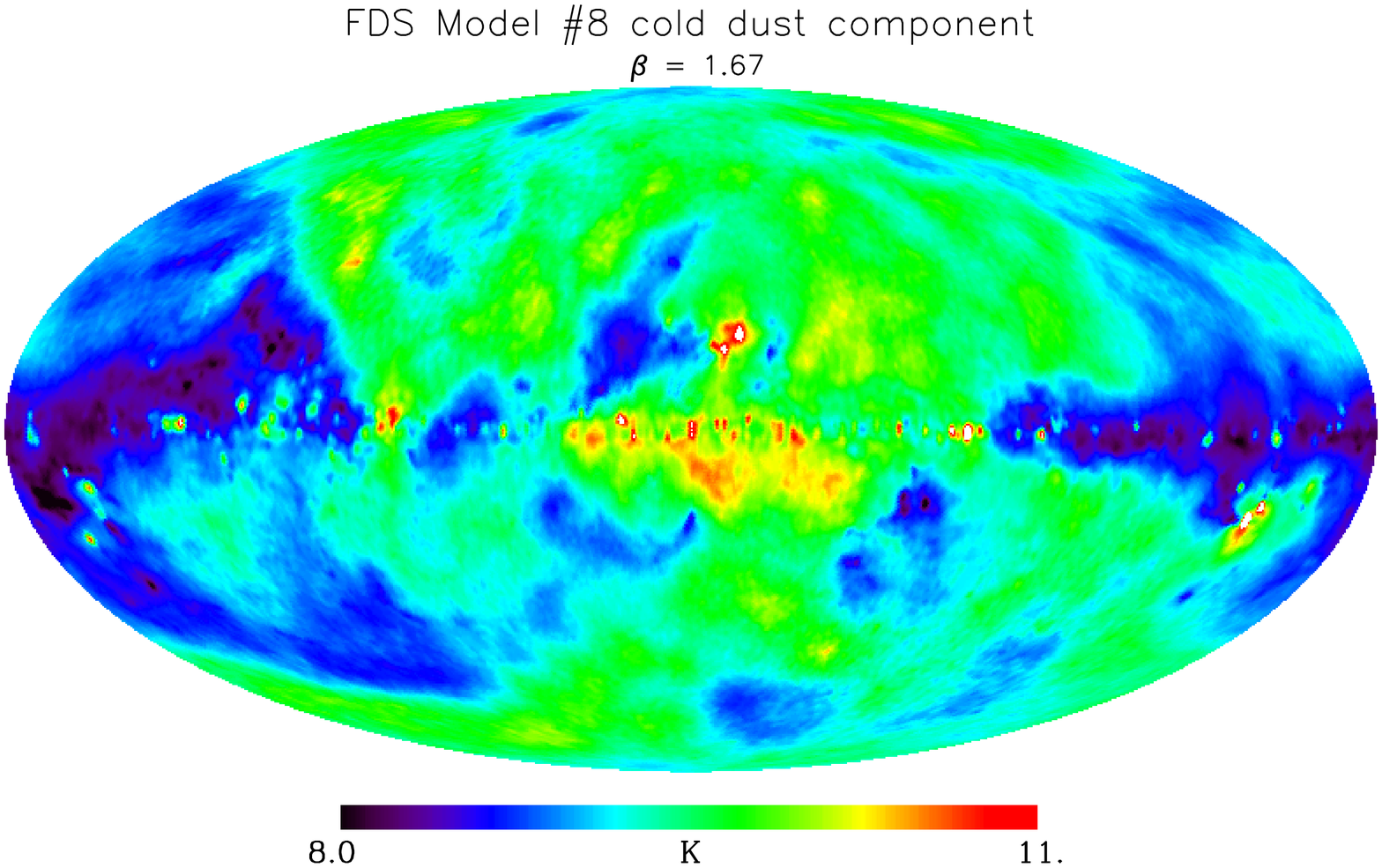}\\
	\end{tabular}
	\hspace{1pc}\\
	\begin{tabular}{c}
      		\includegraphics[scale=0.31]{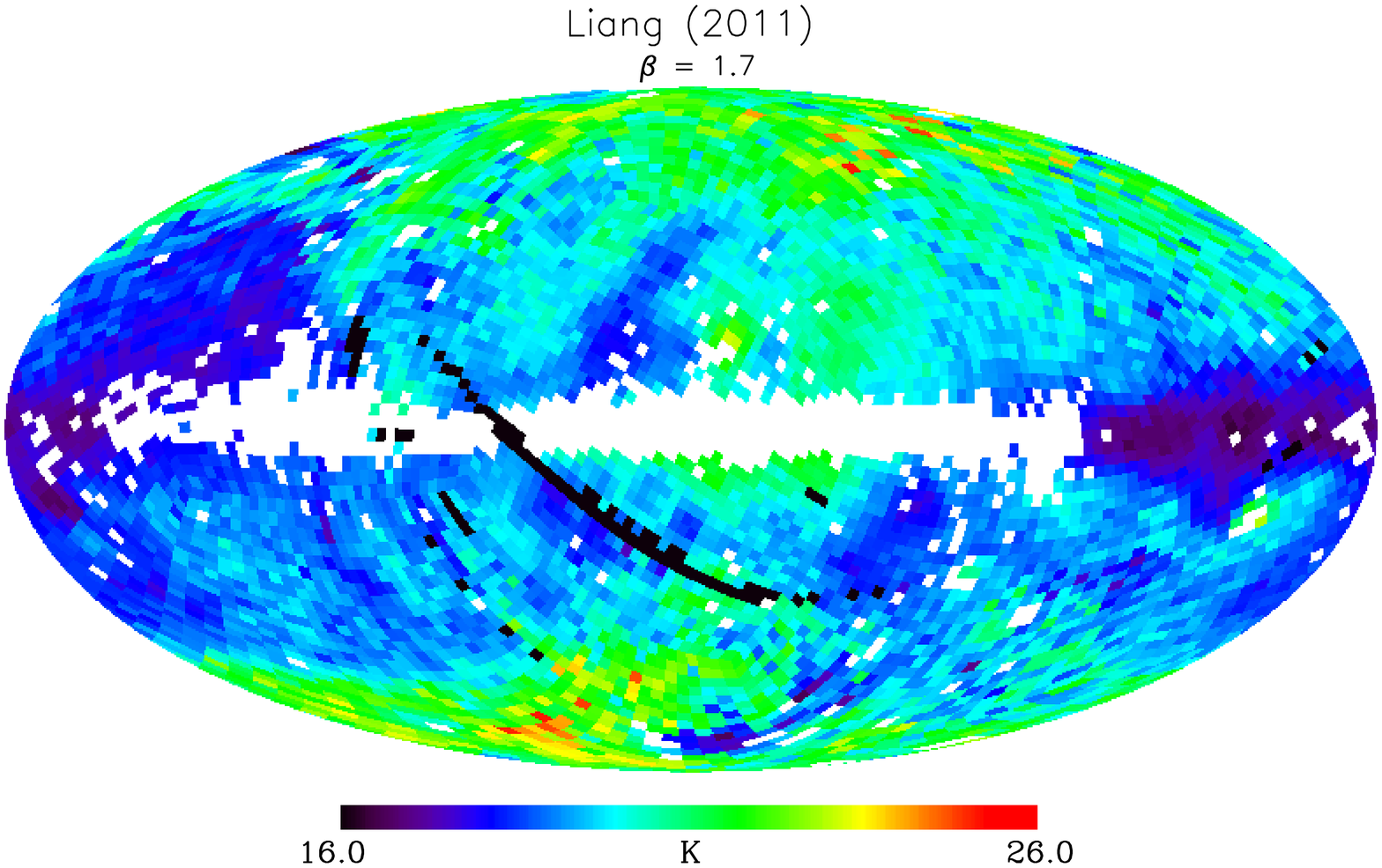}\\
					\hspace{1pc}\\
      		\includegraphics[scale=0.31]{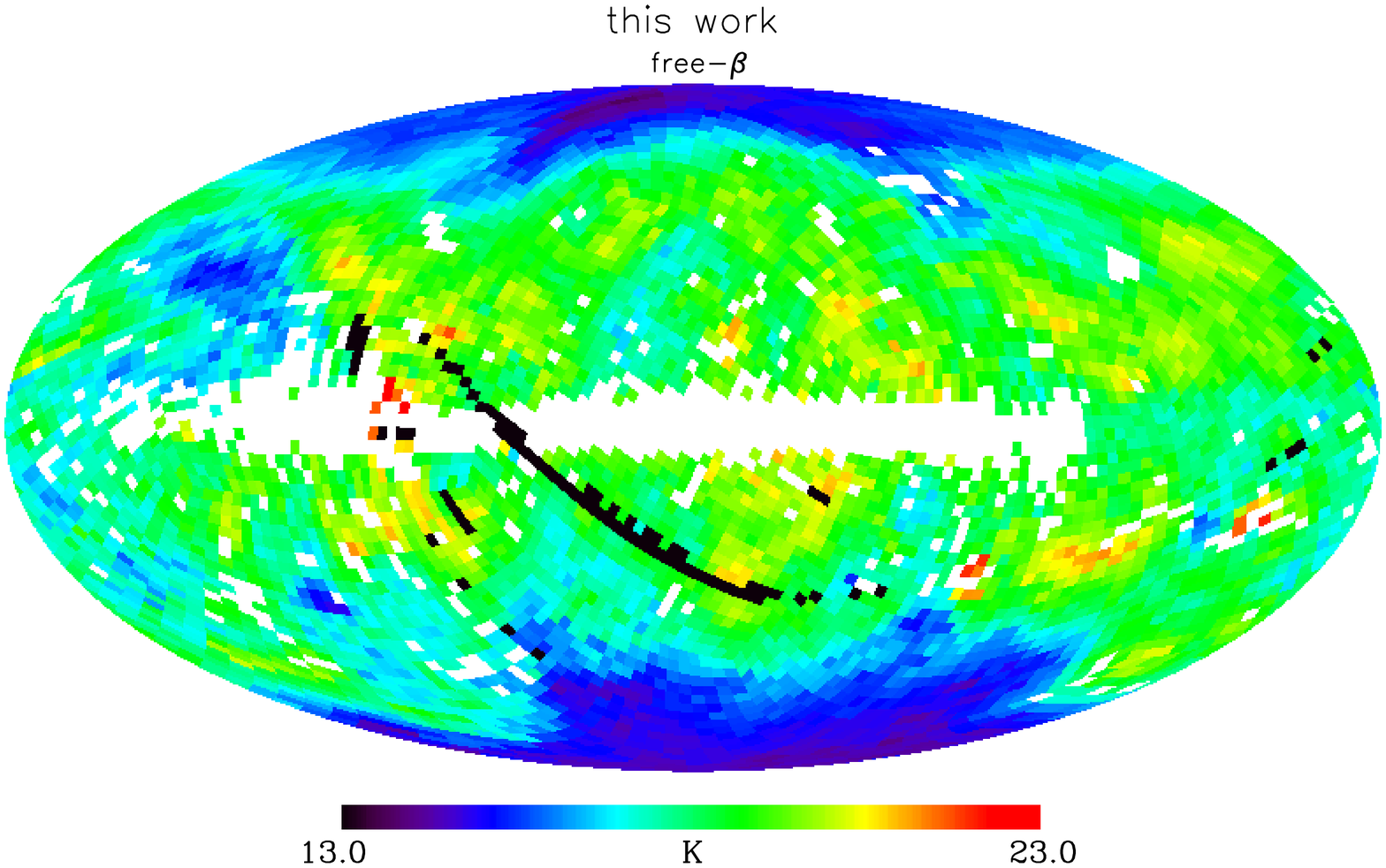}\\
	\end{tabular}
\caption{Dust temperature maps by the two-component $\beta = 2.0$ model of \citet{Reach95} (first row), the two-component $\beta_1 = 2.70$ and $\beta_2 = 1.67$ Model \#8 of \citet{Finkbeiner99} (second row), the one-component $\beta = 1.7$ model with $T_\mathrm{dust} / \delta T_\mathrm{dust} \ge 10$ of \citet{Liang11} (third row), and the one-component free-$\beta$ model with $\beta / \delta \beta \ge 10$ of this work (last row). Data of the Reach model were obtained from their manuscript; the Galactic plane of both maps have been masked  due to incomplete data for the region. Data of the Finkbeiner model were generated from their IDL code ``predict\_thermal.pro.'' Pixels in both the $\beta = 1.7$ and the free-$\beta$ maps are masked in white if they correspond to a $\chi^2_\mathrm{dof} > 1.13$.}
	\label{fig:compare_temp_maps}
\end{figure*}

Fig. \ref{fig:compare_temp_maps} presents dust temperature predictions by the two-component $\beta = 2.0$ model of \citet{Reach95}, the two-component $\beta_1 = 2.70$ and $\beta_2 = 1.67$ Model \#8 of \citet{Finkbeiner99}, the one-component $\beta = 1.7$ model with $T_\mathrm{dust} / \delta T_\mathrm{dust} \ge 10.0$, and the one-component free-$\beta$ model with $\beta / \delta \beta \ge 10.0$. The last two models are the ones derived in Section \ref{sec:results}. We do not have {\sl Planck} Collaboration's dust temperature predictions \citep{Planck11a}, but since their one-component dust model has the emissivity spectral index at 1.8, we use the temperature map of our one-component $\beta = 1.7$ model to approximate their results.

Notice the difference in dust temperature predictions at the Galactic polar caps relative to lower latitude regions in these maps. In the first three cases, where emissivity spectral indices of the models are fixed, temperature estimates around the Galactic polar caps are noticeably higher or at about the same level compared to those at lower latitudes. On the other hand, the free-$\beta$ model predicts the polar caps to have the lowest temperature of the entire Galaxy.

\begin{figure}
	\centering
    \includegraphics[scale=0.54]{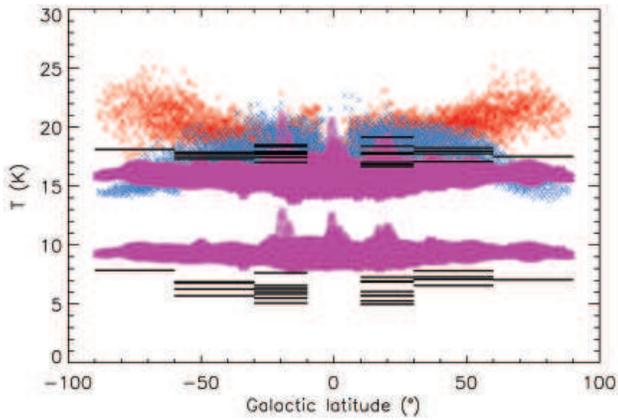}
  \caption{$T_\mathrm{dust}$ vs. Galactic latitude for the two-component $\beta = 2.0$ model of \citet{Reach95} (black horizontal bars), the two-component $\beta_1 = 2.70$ and $\beta_2 = 1.67$ Model \#8 of \citet{Finkbeiner99} (purple), the one-component $\beta = 1.7$ model of \citet{Liang11} (red), and the one-component free-$\beta$ model from this work (blue).}
 	\label{fig:T_vs_glat}
\end{figure}

To perform the comparison in a more quantitative manner, we plot $T_\mathrm{dust}$ vs. Galactic latitude for the different models in Fig. \ref{fig:T_vs_glat}. One may notice that data points corresponding to the one-component $\beta = 1.7$ model flare up at $|b| \ga 30 \degr$ most dramatically among all models. One may also notice that the two temperature components of Model \#8 share the same distributions and that their values are approximately independent of Galactic latitude. In the southern hemisphere, one may notice that the two dust components of the Reach model consistently have higher temperatures at the Galactic polar caps than at mid latitudes. In fact, temperature of the cold dust at the southern polar cap is the highest that cold dust experiences throughout the entire Galaxy. In comparison to these three models, the free-$\beta$ model is the only model which predicts a decrease in dust  temperature with respect to Galactic latitude.

\begin{figure}
	\centering
    \includegraphics[scale=0.55]{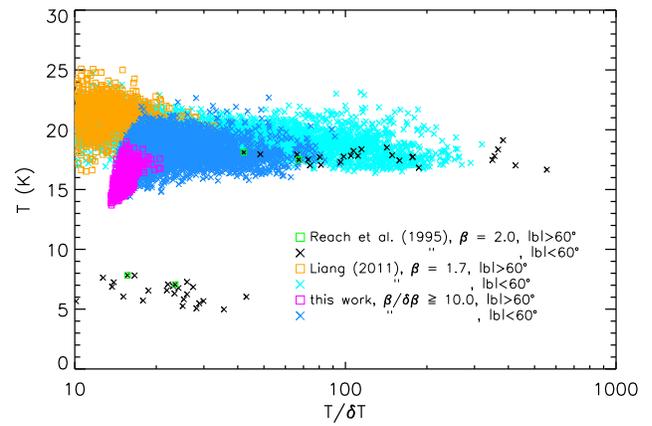}
  \caption{$T_\mathrm{dust}$ vs. $T_\mathrm{dust}/\delta T_\mathrm{dust}$ for the two-component $\beta = 2.0$ model of \citet{Reach95} (green and black), the one-component $\beta=1.7$ model (orange and light blue), and the one-component free-$\beta$ model (magenta and blue). Squares represent fits with $|b| > 60 \degr$, and crosses represent fits with $|b| < 60 \degr$.}
 	\label{fig:T_vs_SN}
\end{figure}

We investigate the possibility that dust temperature predictions are biased positively when signal-to-noise of data decreases. Fig. \ref{fig:T_vs_SN} plots $T_\mathrm{dust}$ as a function of $T_\mathrm{dust}/\delta T_\mathrm{dust}$ as given by the two-component $\beta = 2.0$ model of \citet{Reach95}, the one-component $\beta=1.7$ model, and the one-component free-$\beta$ model. Model \#8 is not included because FDS study does not provide error estimates for their models. Notice that while signal-to-noise of the fits vary by over one, sometimes two, order of magnitude, $T_\mathrm{dust}$ values change by less than 70 per~cent, so $T_\mathrm{dust}$ is approximately independent of the signal-to-noise of the fits. More specifically, within the set of points for each model, in comparison to the (${T_\mathrm{dust}}_{|b| > 60 \degr}$, ${T_\mathrm{dust}/\delta T_\mathrm{dust}}_{|b| > 60 \degr}$) pairs, the existence of lower $T_\mathrm{dust}/\delta T_\mathrm{dust}$ and lower $T_\mathrm{dust}$ pairs or higher $T_\mathrm{dust}/\delta T_\mathrm{dust}$ and higher $T_\mathrm{dust}$ pairs shows that the data do not support a definitive causal relation between low signal-to-noise and high temperature prediction. Based on these observations, we  conclude that any potential bias to the fits due to low signal-to-noise cannot be the main reason for high temperature predictions seen at the Galactic polar caps by fixed-$\beta$ models.

Since we have checked temperature predictions of models with spectral index at the range of most likely values between 1.4 and 2.6 \citep{Liang11}, and they all exhibit higher temperature at the Galactic polar caps than at lower latitudes, we suspect that such a phenomenon is an inherent artifact resulted from fixing the emissivity spectral index at a particular value.
As a result, we argue that the temperature predictions of these fixed-$\beta$ models are unphysical.

In addition to predicting reasonable temperature distribution at the Galactic polar caps, the free-$\beta$ model predicts regions with known heating sources to have higher temperature than their surrounding environment. For instance, we can identify, off the Galactic plane, high temperature regions that correspond to the  Ophiuchus region, $\beta$ Centauri, the Large Magellanic Cloud and the Orion region. At the Galactic plane, model predictions show that dust temperature decreases with increasing Galactocentric distance, consistent with observations of the DIRBE experiment \citep{Sodroski94}. On top of this temperature gradient the temperature map shows the Outer Galaxy to be asymmetrically interspersed with strong infrared emission regions, such as NGC 7538 at $l=111^\circ$, Cas A at $111\fdg7 \la l \la 112\fdg1$, W3/4/5 at $132\fdg5 \la l \la 138\fdg5$, Cygnus region at $75^\circ \la l \la 90^\circ$, and the Gum Nebula at $187\fdg5 \la l \la 193\fdg5$. These regions at the Outer Galaxy are currently masked under the criterion that $\chi^2_{dof}$ of the fit be $\la$ 1.13. For these regions we believe that complex emission sources in addition to thermal dust have contributed to our prepared spectra. This challenges our model's assumption that only thermal dust can contribute to the continuous emission. If true, then our model is no longer sufficient.

\begin{figure}
	\centering
    \includegraphics[scale=0.41]{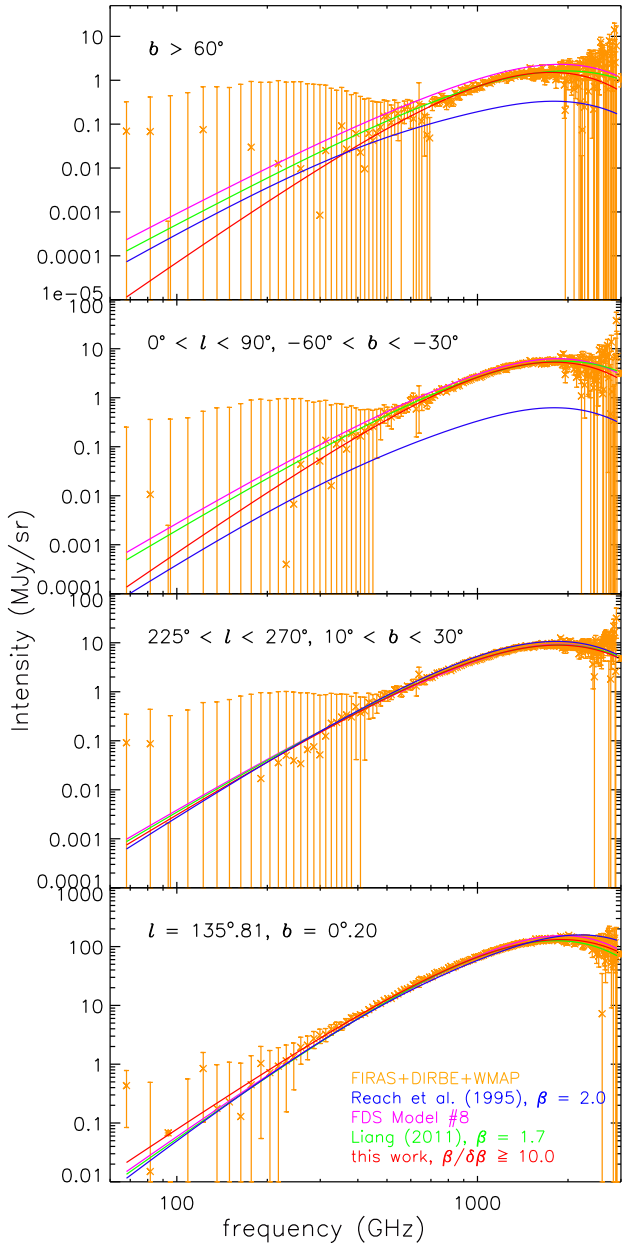}
  \caption{Averaged dust spectra (orange) and predictions of the two-component $\beta = 2.0$ model of \citet{Reach95} (blue), Model \#8 of \citet{Finkbeiner99} (magenta), one-component $\beta = 1.7$ model of \citet{Liang11} (green), and one-component free-$\beta$ model (red).}
 	\label{fig:spectra}
\end{figure}

\begin{figure}
	\centering
    \includegraphics[scale=0.41]{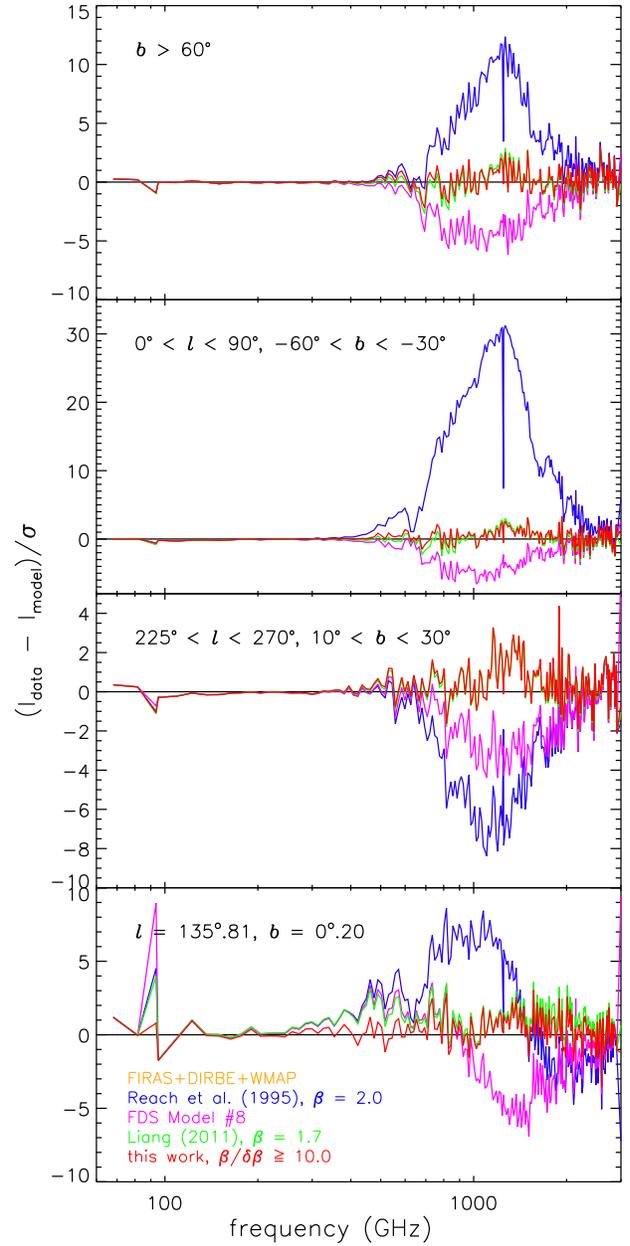}
  \caption{Difference between the averaged 214-channel dust spectra and model predictions. Color-model assignment is the same as in Fig. \ref{fig:spectra}.
The differences are obtained by first subtracting model predictions from each 7$^\circ$ dust spectra,  then dividing the results by measurement uncertainties, and finally averaging the differences.}
 	\label{fig:residuals}
\end{figure}

Fig. \ref{fig:spectra} present samples of fits to the average dust spectra by each of the aforementioned models at different Galactic latitudes. While most models are able to simulate the data with various degrees of success, the free-$\beta$ model is consistently able to trace the data points closely at all Galactic latitudes.
Fig. \ref{fig:residuals} compares the normalized difference between data and intensity predictions for the same set of models and the same sky regions. It shows that the free-$\beta$ model is able to match the data within 1-sigma for majority of the channels while the other models are less consistent at predicting close to measurement values across frequency and Galactic latitude. That the normalized differences of the $\beta = 1.7$ model are comparable to those of the free-$\beta$ model demonstrates the limitation of relying on a single number to evaluate a multifaceted operation, such as using the normalized difference or the $\chi^2$ to differentiate models. Here, the free-$\beta$ model and the $\beta = 1.7$ model have about the same $\chi^2$ value, but one of them has a strong connection with other facts we know of the Galaxy and the other does not.

%% file: discussion_dust6.tex
\subsection{Relation between dust temperature and emissivity spectral index}

The existence of an anti-correlation between the emissivity spectral index and dust temperature has been a subject of debate in the last decade. For example, 
\citet{Veneziani10} did not find a clear inverse correlation between the two parameters in BOOMER{\sevensize AN}G's measurements of the Galactic cirrus. 
\citet{Planck11a} derived all-sky dust temperature using a fixed value of the emissivity spectral index due to degeneracy between temperature and spectral index in a free-$\beta$ fit \citep{Shetty09a, Shetty09b}. And \citet{Planck11c} carried out analysis of dust properties at high Galactic latitudes using the same fixed spectral index model in order to maintain consistency within the Collaboration.

On the other hand, \citet{Dupac01,Dupac02,Dupac03} of the ProNaOS experiment reported evidence of the inverse correlation in their observations of M42, M17 and NGC891, \citet{Desert08} of the Archeops experiment confirmed the correlation on 300 Galactic point sources, \citet{Paradis10} confirmed the existence of the correlation in two Hi-GAL fields at the Galactic plane, and \citet{Planck11b} concluded that observation of objects in the Early Cold Core Catalogue were not consistent with a constant value of the spectral index over all temperature. 

The strongest argument for the existence of a temperature dependent power-law emissivity spectral index comes from laboratory measurements. \citet{Agladze96} measured absorption spectra of crystalline and amorphous silicate grains at 0.7 -- 2.9~mm at 1.2 -- 30~K and found that the wavelength dependent absorption coefficients of the amorphous grains can be modeled with power laws, and the exponents of which vary with temperature. Agladze et al. attributed the correlation to a two-level population effect found in the low-lying tunneling states in bulk glasses. \citet{Mennella98} measured the absorption coefficients of crystalline and amorphous carbon and silicate grains at 20~$\mu$m -- 2~mm at 4 -- 295~K and found that the absorption coefficient followed a power law with the spectral index depending on grain temperature. They attributed the observed pattern to the dominance of two-phonon difference processes. In 2001, Chihara et al. measured far-infrared absorption spectra of crystalline circumstellar dust analogues between 20 -- 80~$\mu$m at 4.2~K and at 295~K. They also reported a higher value of the spectral index at 4.2~K than at 295~K. In 2005, Boudet et al. measured the mass absorption coefficients of amorphous silicates between 100 $\mu$m -- 2~mm at 10 -- 30~K and found that the power law spectral index had a strong temperature and wavelength dependence. More specifically, the anti-correlation between spectral index and grain temperature was more pronounced at 500~$\mu$m -- 1~mm than at 100 -- 250~$\mu$m. Comparing their measurements with the temperature-spectral index relation deduced from synthesis of the ProNaSO observations, Boudet et al. found a comparable anti-correlation between temperature and spectral index between observations and laboratory measurements when the spectral index is between 1 and 2.7. They attributed the origin of the relation to the tunneling processes in two level systems. Bringing together evidence from laboratory measurements and astronomical observations, \citet{Meny07} proposed a model that relates the emission spectrum's dependence on temperature and internal structure of the grains. 
Additional evidence in support of the explanation that internal grain structures give rise to the apparent correlation between dust temperature and spectral index came in when \citet{Coupeaud11} reported that mass absorption coefficients of synthesized analogues of interstellar amorphous silicate grains decreased with temperature and that the local spectral index anti-correlated with grain temperature. These laboratory results strengthen our confidence that the anti-correlation between dust temperature and spectral index seen in astronomical observations is not spurious but instead grounded on the physical properties of interstellar dust grains.

\begin{figure}
	\centering
     	\includegraphics[scale=0.55]{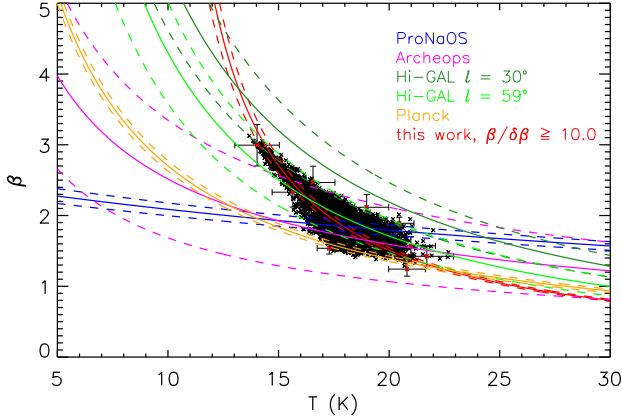}
	\caption{Data and the best-fitting $\beta-T_\mathrm{dust}$ relations from this and other observations. The black crosses and error bars represent best-fitting values from one-component free-$\beta$ fits that satisfy $\beta /\delta \beta \ge$ 10.0. Both data and errors are used in fitting the hyperbolic function $\beta = 1/ (\delta+\omega \cdot T_\mathrm{dust})$. Only fits with $\chi^2_\mathrm{dof} \le 1.13$ are included. 
Also presented are the best-fitting $\beta-T_\mathrm{dust}$ relation of the ProNaOS experiment (blue, \citealt{Dupac03}), the Archeops experiment (magenta, \citealt{Desert08}), the two Hi-GAL fields (green, \citealt{Paradis10}), and the {\sl Planck} mission (orange, \citealt{Planck11b}). Dashed curves represent 1-sigma uncertainties of the respective models.}
	\label{fig:alpha_vs_T_free_alpha_var_res}
\end{figure}

\begin{figure}
	\centering
      	\includegraphics[scale=0.55]{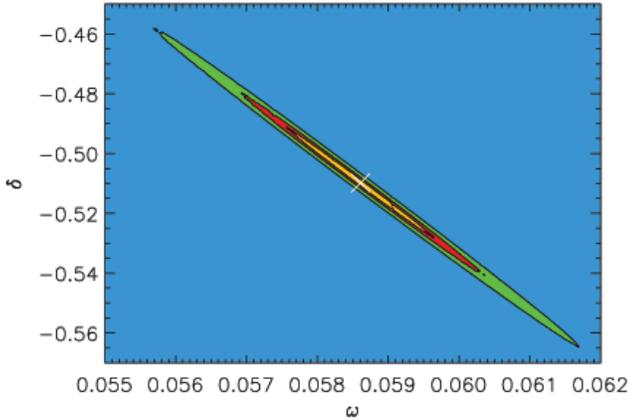}
	\caption{Constant $\chi^2$ contours corresponding to confidence intervals with 68, 95 and 99.99~per~cent probability. We fit the relation $\beta = 1/ (\delta+\omega \cdot T_\mathrm{dust})$ to the best-fitting $T_\mathrm{dust}$ and $\beta$ values from the one-component free-$\beta$ model. The white cross represents location of the minimum $\chi^2$.} 
	\label{fig:beta-T_inverse_correlation_coeff}
\end{figure}

To test the possibility of an inverse correlation between emissivity spectral index and dust temperature in our data, we fit the best-fitting $\beta$ and $T_\mathrm{dust}$ values of the free-$\beta$ model with  $\beta / \delta \beta \ge 10.0$ to the hyperbolic function $\beta = 1/(\delta+\omega \cdot T_\mathrm{dust})$ and obtain $\delta = -0.510 \pm 0.011$ and $\omega = 0.059 \pm 0.001$ with a $\chi^2_\mathrm{dof} = 0.99$. 
Fig. \ref{fig:alpha_vs_T_free_alpha_var_res} presents the data and the $T_\mathrm{dust}$-$\beta$ relations identified by this and other works. Notice that our best-fitting $T_\mathrm{dust}$-$\beta$ relation overlaps with all but that of the Hi-GAL field at $l = 30\degr$. Since the ($T_\mathrm{dust}$, $\beta$) pairs we used to constrain the hyperbolic function come from all over the sky except for about 13 per~cent of the total sky area in the Inner Galaxy, the overlap with other experiments' results confirms that our $T_\mathrm{dust}$-$\beta$ correlation has a broad representation. That it is within one-sigma of the $T_\mathrm{dust}$-$\beta$ relation identified in \citet{Paradis10} for the $l=30\degr$ field may have to do with one or both of the following reasons. First, the physical properties of dust may have changed from the Inner Galaxy to regions far away from active star formation. If so, then our set of ($T_\mathrm{dust}$, $\beta$) values is not a good representation of dust properties in the $l=30\degr$ field since fits of the Inner Galactic spectra are not included. Secondly, one may understand the isolation of the curve for Hi-GAL field at $l=30\degr$ from explanation given in \citet{Paradis10}, which points to the fact that data from the $l=30\degr$ field challenge their model's single-temperature assumption. Finally, that our set of ($T_\mathrm{dust}$, $\beta$) values marginally overlap with the relation from the {\sl Planck} study of cold cores in the Galaxy \citep{Planck11b} may have to do with the fact that there is only a small overlap between the two data sets:  
While the 7608 cold cores in the \citet{Planck11b} analysis largely distribute along the Galactic plane, our model's target regions are at mid and high latitudes and part of the Outer Galaxy. In addition, the 7608 cold cores form a selective group of objects. To be selected, an object has to be a compact source colder than the surrounding envelope and a 17-K diffuse Galactic background. This last criterion would have excluded over 73 per cent of our data points. Fig. \ref{fig:beta-T_inverse_correlation_coeff} presents the constant $\chi^2$ contours, corresponding to confidence intervals with 68, 95 and 99.99 per cent probability. It shows that a temperature independent $\beta$ model is rejected by the data with more than 99.99~per~cent probability, an indication for a real dependence between $T_\mathrm{dust}$ and $\beta$.

We present a summary of the best-fitting $\delta$ and $\omega$ for other levels of constraint on the fits in Table \ref{tbl:alpha_T}. The exact functional form of the temperature-dependent spectral index cannot be established by our data alone. However, that a physically sensible fit to the all-sky data requires a free-$\beta$ model and that an inverse relation fits the $\beta$ and $T_\mathrm{dust}$ values well give strong support to the anti-correlation between the emissivity spectral index and dust temperature.

\begin{table*}
\begin{minipage}{166mm}
\centering
\caption[center]{Best-fitting parameters to the relation $\beta = 1/(\delta+\omega \cdot T_\mathrm{dust})$ between emissivity spectral index and dust temperature}
\begin{tabular}{|c|c|c|c|c|} 
\hline
constraint on fits & $\delta$ & $\omega$ & number of data points & $\chi^2_\mathrm{dof}$ \\ 
\hline
no constraint & $-0.305 \pm 0.014$ & $0.048 \pm 0.001$ & 5421 & 0.72 \\		
$\beta /\delta \beta \ge ~5.0$  & $-0.399 \pm 0.013$ & $0.053 \pm 0.001$ & 5386 & 0.83 \\		%
$\beta /\delta \beta \ge ~6.7$ & $-0.441 \pm 0.012$ & $0.055 \pm 0.001$ & 5357 & 0.88 \\
$\beta /\delta \beta \ge 10.0$ & $-0.510 \pm 0.011$ & $0.059 \pm 0.001$ & 5305 & 0.99 \\
\hline
\end{tabular}
\label{tbl:alpha_T}\\
\end{minipage}
\end{table*}

\begin{figure}
	\centering
      	\includegraphics[scale=0.55]{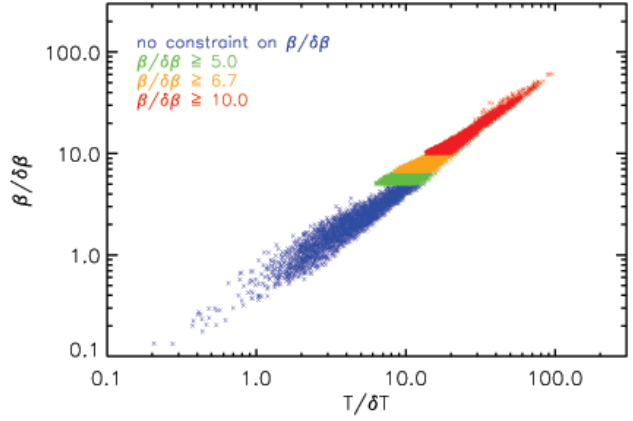}
	\caption{Scatter plot of $\beta / \delta \beta$ and $T_\mathrm{dust} / \delta T_\mathrm{dust}$. The blue, green, yellow and red data points represent values of $\beta / \delta \beta$ and $T_\mathrm{dust} / \delta T_\mathrm{dust}$ from fits that use no constraint on any parameter and those that satisfy $\beta /\delta \beta \ge$ 5.0, 6.7 and 10.0, respectively. Only fits with $\chi^2_\mathrm{dof} \le 1.13$ are included.} 
	\label{fig:dela_vs_delT_free_alpha}
\end{figure}	

The observed correlation between emissivity spectral index and dust temperature in our data is likely to be intrinsic rather than from noise. In 2009, Shetty et al. showed that the accuracy of parameter estimation depends on the wavelength range of the spectra used in model fitting. They show that for dust temperature $\le 20$~K, fits using spectra at $100-600~\mu$m are not sensitive to noise. However, if 40 K $\le T_\mathrm{dust} \le 80$~K, a 5~per~cent noise in a data set of different $T_\mathrm{dust}$ and the same $\beta$ may generate a pattern in the best-fitting $\beta$ vs. $T_\mathrm{dust}$ plot that can be confused with a true anti-correlation in the two parameters. Since our spectra, coming at $100-4400~\mu$m, well sample the Rayleigh-Jeans tail of the peak of the $\sim 20$~K modified blackbody spectrum, our best-fitting parameters are good approximates of the true values.

As the quality of data improves, the free-$\beta$ model will allow future studies to pinpoint the values of $\beta$ and $T_\mathrm{dust}$ with greater accuracy. In Fig. \ref{fig:dela_vs_delT_free_alpha} we present $\beta / \delta \beta$ vs. $T_\mathrm{dust} / \delta T_\mathrm{dust}$ of the all-sky collections of fits with different amounts of constraint on $\beta / \delta \beta$. It shows that the uncertainty in $\beta$ and $T_\mathrm{dust}$ are directly proportional while $\beta$ is more susceptible to errors than $T_\mathrm{dust}$. That is, with more sensitive data one will be able to reduce uncertainty in $\beta$ and $T_\mathrm{dust}$ at the same time.

%% file: conclusion_dust6.tex
We have examined models of interstellar dust thermal emission at far-infrared and millimeter wavelengths. 
Starting with deducing a new and improved version of dust spectra from FIRAS calibrated measurements and unifying them with measurements of the DIRBE and the {\sl WMAP}, one-component dust models with fixed and variable emissivity spectral index are fit to 6063 214-channel spectra at fixed and variable spatial resolutions.

We show that the free-$\beta$ model can predict more physically motivated dust temperature than models with fixed emissivity spectral index. 
With observations showing that the intensity of dust emission and the amount of dust decrease with the increase in Galactic latitude and that fewer and weaker stars exist at higher Galactic latitudes, dust temperature is expected to be lower at the Galactic polar caps than at lower latitudes. Among all-sky dust models found in the literature and the one-component dust models constructed in our study, we demonstrate that only the free-$\beta$ model passes this test. All fixed-$\beta$ models predict higher or the same temperatures at the Galactic polar caps as at lower latitudes.

Our best dust model is the one-component free-$\beta$ model with $\beta/\delta \beta \ge 10.0$. It fits dust spectra over 86~per~cent area of the full sky. Dust temperature is predicted to be $13.69-22.69$~($\pm 1.26$)~K, emissivity spectral index to be $1.2-3.1$~($\pm 0.31$), and optical depth to range $0.61-46.15 \times 10^{-5}$ at $\nu_0 = 900$~GHz ($\lambda_0 = 333 \mu$m) with a 23.25~per~cent uncertainty. Our model has the 7$^\circ$ angular resolution of FIRAS and uses variable spatial averaging at high latitudes. 
It supports the interpretation of an anti-correlation between emissivity spectral index and dust temperature. We fit the relation $\beta = 1/(\delta+\omega \cdot T_\mathrm{dust})$ to the $\beta$ and $T$ values and obtained $\delta = -0.510 \pm 0.011$ and $\omega = 0.059 \pm 0.001$. A temperature independent $\beta$ model ($\omega=0$) is rejected by the data with more than 99.99~per~cent probability, an indication that a real dependence between $T_\mathrm{dust}$ and $\beta$.

The free-$\beta$ model can be used to remove dust contamination in CMB measurements. It can also serve as an all-sky extended-frequency-range reference for other experiments. With higher resolution data sets, such as those from {\sl Planck}, one will be able to further refine our knowledge of interstellar dust grains and their interaction with other constituents in the Galaxy.

%% file: bibliography_dust6.tex
%\addcontentsline{toc}{chapter}{Bibliography}

%\bibliography{References}